\documentclass[pdftex,letterpaper,12pt,longtitle,authoryear,final]{elsarticle}
\usepackage{amssymb}
\usepackage{longtable}
\usepackage{lscape}
\usepackage[switch]{lineno}
\usepackage{array}

\usepackage{textcomp}

\usepackage{fancyhdr}
\bibpunct[ ]{(}{)}{,}{a}{}{,}

\usepackage[width=6.5in,height=9in]{geometry}
\usepackage{hyperref}

\newif\ifincludegraphics
\includegraphicstrue

\newcommand\afrho{\mbox{$Af\rho$}}
\newcommand\efrho{\mbox{$\epsilon f \rho$}}
\newcommand\micron{\mbox{\textmu{}m}}%
\newcommand\arcdeg{\mbox{$^\circ$}}%
\newcommand\arcmin{\mbox{$^\prime$}}%
\newcommand\arcsec{\mbox{$^{\prime\prime}$}}%
\newcommand\degr{\arcdeg}%
\newcommand\mjysr{MJy~sr$^{-1}$}

\newcommand\sun{\odot}%
\newcommand\ten[1]{\mbox{$\times10^{#1}$}}
\newcommand\inv{$^{-1}$}

\newcommand\coo{CO$_2$}
\newcommand\water{H$_2$O}
\newcommand\sst{\textit{Spitzer Space Telescope}}
\newcommand\spitzer{\textit{Spitzer}}

\newcommand\rosetta{\textit{Rosetta}}
\newcommand\di{\textit{Deep Impact}}

\newcommand\nodata{ ~$\cdots$~ }%

\defcitealias{fernandez11}{Paper~I}

\journal{Icarus}

\begin{document}

\begin{frontmatter}

\title{The Persistent Activity of Jupiter-Family Comets at 3 to 7 AU}

\author[umd,ucf]{Michael S. Kelley\corref{cor}}
\cortext[cor]{Corresponding author}
\ead{msk@astro.umd.edu}

\author[ucf]{Yanga R. Fern\'andez}
\ead{yfernandez@physics.ucf.edu}

\author[iac]{Javier Licandro}

\author[apl]{Carey M. Lisse}

\author[sofia]{William T. Reach}

\author[umd]{Michael F. A'Hearn}

\author[jpl]{James Bauer}

\author[ucf]{Humberto Campins}

\author[qub]{Alan Fitzsimmons}

\author[lam]{Olivier Groussin}

\author[lam]{Philippe L. Lamy}

\author[kent]{Stephen C. Lowry}

\author[uh]{Karen J. Meech}

\author[uh]{Jana Pittichov\'a}

\author[mps]{Colin Snodgrass}

\author[ko]{Imre Toth}

\author[apl]{Harold A. Weaver}

\address[umd]{Department of Astronomy, University of Maryland, College
  Park, MD 20742-2421, USA}

\address[ucf]{Department of Physics, University of Central Florida,
  4000 Central Florida Blvd., Orlando, FL 32816-2385, USA}

\address[iac]{Instituto de Astrof\'isica de Canarias, c/V\'ia
  L\'actea s/n, 38205, La Laguna, Tenerife, Spain}

\address[apl]{Applied Physics Laboratory, Johns Hopkins University,
  11100 Johns Hopkins Rd, Laurel, MD 20723, USA}

\address[sofia]{Stratospheric Observatory for Infrared Astronomy,
  Universities Space Research Association, MS 211-3, Moffett Field, CA
  94035, USA}

\address[jpl]{NASA Jet Propulsion Laboratory, 4800 Oak Grove Dr.,
  Pasadena, CA 91109, USA}

\address[qub]{Astrophysics Research Centre, School of Physics and
  Astronomy, Queen's University Belfast, Belfast BT7 1NN, UK}

\address[lam]{Aix Marseille Universit\'e, CNRS, LAM (Laboratoire
  d'Astrophysique de Marseille) UMR 7326, 13388, Marseille, France}

\address[kent]{Centre for Astrophysics and Planetary Science,
  University of Kent, Ingram Building, Canterbury, Kent CT2 7NH, UK}

\address[uh]{Institute for Astronomy, University of Hawaii, 2680
  Woodlawn Drive, Honolulu, HI 96822, USA}

\address[mps]{Max-Planck-Institut f\"ur Sonnensystemforschung,
  Max-Planck-Str. 2, 37191 Katlenburg-Lindau, Germany}

\address[ko]{Konkoly Observatory, PO Box 67, Budapest 1525,
  Hungary}

\begin{abstract}
We present an analysis of comet activity based on the \sst{} component
of the Survey of the Ensemble Physical Properties of Cometary Nuclei.
We show that the survey is well suited to measuring the activity of
Jupiter-family comets at 3--7~AU from the Sun.  Dust was detected in
33 of 89 targets ($37\pm6$\%), and we conclude that 21 comets
($24\pm5$\%) have morphologies that suggest ongoing or recent cometary
activity.  Our dust detections are sensitivity limited, therefore our
measured activity rate is necessarily a lower limit.  All comets with
small perihelion distances ($q < 1.8$~AU) are inactive in our survey,
and the active comets in our sample are strongly biased to
post-perihelion epochs.  We introduce the quantity \efrho{}, intended
to be a thermal emission counterpart to the often reported \afrho, and
find that the comets with large perihelion distances likely have
greater dust production rates than other comets in our survey at
3--7\,AU from the Sun, indicating a bias in the discovered
Jupiter-family comet population.  By examining the orbital history of
our survey sample, we suggest that comets perturbed to smaller
perihelion distances in the past 150~yr are more likely to be active,
but more study on this effect is needed.

\end{abstract}

\end{frontmatter}
\thispagestyle{fancy}

\section{Introduction}

The Survey of the Ensemble Physical Properties of Cometary Nuclei
(SEPPCoN) provides a data set with which we may examine comet activity
at intermediate heliocentric distances, $r_h$ (here, 3--7~AU).  The
primary goal of SEPPCoN is to measure, in a consistent manner, the
sizes and surface properties of a statistically significant number
($\approx100$) of Jupiter-family comet nuclei.  SEPPCoN is comprised
of two observational campaigns: a mid-infrared (mid-IR) component
utilizing the imaging capabilities of the \sst{} \citep{werner04}, and
a visible light component using ground-based instrumentation.  Early
results from the \spitzer{} survey on Comets 22P/Kopff and
107P/Wilson-Harrington were presented by \citet{groussin09} and
\citet{licandro09}, respectively.  \citet{fernandez11}, hereafter
\citetalias{fernandez11}, presented the first survey results from
SEPPCoN.  In \citetalias{fernandez11}, we measured and analyzed the
thermal emission from 89 comet nuclei, and presented results on the
cumulative size distribution and spectral properties of the \spitzer{}
survey targets.  The goal of the present paper is to examine the dust
and activity of those same targets.

Active comets produce a coma.  A coma is a gravitationally unbound
atmosphere, driven by the outflow of volatiles from the nucleus
surface or sub-surface.  As comets approach the Sun, their surfaces
warm, eventually causing gases to be released from the nucleus due to
sublimation of volatile ices.  The heliocentric distance at which this
occurs depends on the physical properties of the nucleus in question:
shape, composition, and internal structure.  Volatile driven mass loss
due to insolation is typical of comet activity, i.e., we may not even
consider an object to have \textit{cometary activity} unless we
observe a coma and/or tail generated by solar heating of volatiles (as
opposed to impact driven mass loss).  In addition to the gases, the
coma typically includes escaping dust and/or icy grains lifted off the
nucleus by the gas outflow.  As a comet recedes from the Sun, the
nucleus cools and activity may be quenched.

Water ice is typically the primary driver of comet atmospheres inside
of 3~AU \citep{meech04-activity}.  Water is by far the most abundant
ice near the nucleus surface, as inferred from remote
spectrophotometric compositional studies of comae.  The next most
abundant gases are \coo{} and CO with relative fractions $\lessapprox
20$\% \citep{bockelee-morvan04, ootsubo12}.  However, \coo{} can also
drive activity, even though it is generally less abundant than water,
as shown by the \textit{Deep Impact} flybys of Comets 103P/Hartley~2
\citep{ahearn11} and 9P/Tempel~1 \citep{feaga07}.  Outside of 3~AU,
the relative contributions of water ice and more volatile ices to
activity is less well understood.  Activity has been observed in more
than 80 comets with large perihelion distances ($q>5$~AU), despite the
inferred low surface temperatures at such great distances.  Ices more
volatile than water must play larger roles in driving activity for
these distant comets than for comets closer to the Sun.  Water
sublimates at $T\approx170$~K (which occurs in the solar system near
$r_h\approx3$~AU), whereas \coo{} ice sublimates at $T\approx80$~K
($r_h\approx12$~AU) and CO at $T\approx20$~K ($r_h\approx50$~AU).
Recent results from the \textit{Akari} satellite show that the coma
mixing ratio of \coo{} to \water{} is systematically larger for comets
outside of 2.5~AU, indicating the diminishing role of water
sublimation as heliocentric distance increases \citep{ootsubo12}.  In
addition to sublimation of ices, the crystallization of amorphous
water ice could release trapped gases and drive activity at 120--160~K
\citep{schmitt89, meech04-activity}, and has been proposed as the
dominant driver of activity in Centaurs \citep{jewitt09-centaurs}.
Whatever the mechanism that drives activity in comets with $q>5$~AU,
not all comets are active at such large heliocentric distances.  By
studying the activity of comets at intermediate heliocentric distances
we may learn which properties of comet nuclei initiate activity as
they approach the Sun, and quench or sustain activity as they recede
from the Sun.

Mid-IR broadband images of comets, such as those taken as part of our
survey, are dominated by thermal emission from the nuclei and
surrounding dust.  With few exceptions, the gas is undetected or only
forms a minor component of the emission.  Because gas expansion is a
critical component to comet activity, but remains undetected in most
mid-IR images, in this work we infer activity from the presence and
morphology of dust structures larger than the point source nucleus,
rather than from the direct detection of gases.

With SEPPCoN \spitzer{} images, we measure the dust activity of comet
nuclei at 3--7~AU, distances where water ice sublimation is typically
low.  First, we review the \spitzer{} observations which were
presented in detail in \citetalias{fernandez11} (\S\ref{sec:obs}).
Next, we present the morphology and the photometric properties of the
dust detected in the survey, and assess the nature of the dust
(\S\ref{sec:results}).  Then, we discuss the frequency of cometary
activity versus heliocentric distance and other parameters, their
implications on the structure and heating of comet nuclei, and analyze
the color temperature of comet dust at 3--7~AU
(\S\ref{sec:discussion}).  Finally, we summarize our findings
(\S\ref{sec:conclusions}).

\section{Observations and Reduction}\label{sec:obs}
The purpose of the \spitzer{} component of the survey is to obtain a
robust estimate of the size distribution of known Jupiter-family comet
nuclei.  To this end, 100 JFCs were selected that met the following
criteria: 1) the ephemeris was constrained well enough such that the
comet could be expected to lie within a $5\arcmin\times5\arcmin$ field
of view (the footprint of \spitzer's largest arrays); 2) the comet
must have been observable by \spitzer{} and beyond $\approx4$~AU from
the Sun during \spitzer{} Cycle 3 (July 2006 to July 2007); and, 3)
the nucleus must have been brighter than $V=24.0$~mag to make
ground-based optical observations feasible.  Criterion 3 requires an
estimate of the nucleus radius ($R$), which we obtained from the
compilation of \citet{lamy04}.  If no estimate existed, we used the
following assumption: $R=1.0$~km for comets with perihelion distances
$q<2.0$~AU, $R=1.5$~km for $2.0<q<2.5$~AU, and $R=2.0$~km for
$q>2.5$~AU.  The assumption attempts to account for the fact that it
is increasingly difficult to discover smaller comets at larger
perihelion distances.

Two \spitzer{} instruments were well suited for the survey: the
24-\micron{} ($\lambda_{eff}=23.7$~\micron, 2.55~arcsec~pixel\inv,
$5.4\arcmin\times5.4\arcmin$ field of view) camera of the Multiband
Imaging Photometer for Spitzer \citep[MIPS;][]{rieke04}, and the
16-\micron{} and 22-\micron{} Infrared Spectrograph
\citep[IRS;][]{houck04} peak-up arrays ($\lambda_{eff}=15.8$ and
22.3~\micron, 1.85~arcsec~pixel\inv, $0.9\arcmin\times1.4\arcmin$
field of view).  Our choice of instrument was based on each comet's
ephemeris uncertainties.  If the $3\sigma$ uncertainty was under
30\arcsec{} we selected the IRS peak-up arrays; if the uncertainty was
between 30 and 200\arcsec{} we selected the 24-\micron{} MIPS camera.
The two IRS peak-up arrays were preferred because they allowed us to
measure a color for each nucleus, which provides an indication of the
effective temperature of the surface (required for the size estimate).

The \spitzer{} spacecraft tracked each comet given the computed
ephemerides, and observed each target twice.  For the IRS, each target
was first observed with the 16-\micron{} peak-up array, followed
immediately with the 22-\micron{} peak-up array.  IRS peak up
background observations were obtained in the array that was not
centered on the comet, although the background observations were not
always useful due to variable detector artifacts (especially latent
charges), and nearby background sources (stars).  For the MIPS, a
duplicate (shadow) observation was executed 1 to 30~hr after the
primary observation.  The shadow observation was close enough in time
to the primary such that the comet remained within the MIPS field of
view but displaced from the original position.  The goal
signal-to-noise ratio on the total flux of each nucleus observation
was 30, based on the radius estimates discussed above.  In
Table~\ref{tab:targets} we list the 89 comets identified in
\citetalias{fernandez11} with their observing circumstances and
relevant orbital parameters.  Details on the general image reduction,
and the identification of specific targets are presented in
\citetalias{fernandez11}.  Out of the 100 targeted comets, two were
not in the field of view of their observations, six were in the field
of view but were not detected, and three have an unknown status (i.e.,
the ephemeris is sufficiently uncertain that we cannot conclude if the
comet was in the field of view).

\section{Results}\label{sec:results}
\subsection{Dust morphology}\label{sec:morphology}
There are three types of comet dust morphologies relevant to this
paper: comae, tails, and trails.  The coma begins at the surface of
the nucleus.  In the rest frame of the nucleus, entrained dust grains
move away from the surface.  At a telescope we observe a roughly
elliptically shaped coma with a surface brightness distribution that
decreases with distance from a central source.  The presence of this
extended dust coma is the best evidence for recent and ongoing
activity in mid-IR observations.

As the coma expands, the morphology becomes increasingly dependent on
the dynamical and physical properties of the dust.  Gas expansion
accelerates dust grains from the nucleus surface and places each grain
into a new orbit around the Sun.  The grain's orbit depends on its
velocity and response to solar radiation pressure.  The radiation
force is size dependent.  In general, smaller grains feel a greater
acceleration from radiation pressure.  The trend reverses in the
sub-micron range where grains are too small to efficiently absorb
solar radiation \citep{burns79}.  In studies of comet grain dynamics,
the radiation force is commonly expressed as the unitless parameter
$\beta = F_r / F_g \propto Q_{pr}/a$, where $F_r$ is the force from
solar radiation, $F_g$ is the force from solar gravity, $Q_{pr}$ is
the grain-dependent radiation pressure efficiency, and $a$ is the
grain radius.

The nucleus may also experience non-gravitational forces.  The comet
nucleus does not feel an appreciable radiation force, but instead
insolation drives mass loss that produces significant secular changes
in the comet's orbit \citep{whipple50-encke, marsden73}.  Altogether,
the nucleus and the dust grains are in different heliocentric orbits,
and as the dust coma expands it transforms into a dust tail.  The
presence of a dust tail may indicate recent activity, but it is not as
conclusive as the presence of a dust coma.  Ambiguity is present
because the tail, as projected on the sky, may be composed of
intermediate-sized slow-moving grains that were ejected weeks prior to
the observation, whereas smaller grains will have much shorter
lifetimes.

The largest grains are removed from the nucleus with the lowest
ejection velocities, and weakly interact with solar radiation (i.e.,
they have small $\beta$ values).  The heliocentric orbits of the large
grains are so similar to the nucleus that only months or years after
ejection will their presence be apparent as a long, linear dust
feature along or near the projected orbit of the nucleus
\citep{sykes92}.  Such a linear structure may be interpreted as a dust
trail, and its presence indicates activity on month-to-year
timescales.

For each image, we searched for the presence of dust using two
methods: PSF matching and visual inspection.  In
\citetalias{fernandez11}, we fitted the central source of each comet
with a scaled PSF.  Those comets with residual emission in excess of a
smooth background were identified as potentially having dust.  Visual
inspection of each image served to verify the initial results from the
PSF fitting, and to identify dust emission outside of the PSF fit
radius (typically 5 to 8 pixels).  The images were inspected with a
range of color scales and smoothing techniques to verify potential
dust.  Since there are two images of each comet, finding dust in both
images gives credence to our interpretation, but is not strictly
required as contamination from artifacts and background sources could
obscure dust in one image but not the other.  In practice, dust is
easily found via visual inspection for most targets.  Only in images
of comet 50P/Arend were the PSF fit residuals suggestive of dust that
could not be verified with visual inspection due to confusion with
background sources (star streaks).

\begin{figure}
  \ifincludegraphics
  \begin{center}
    \includegraphics[width=0.83\columnwidth]{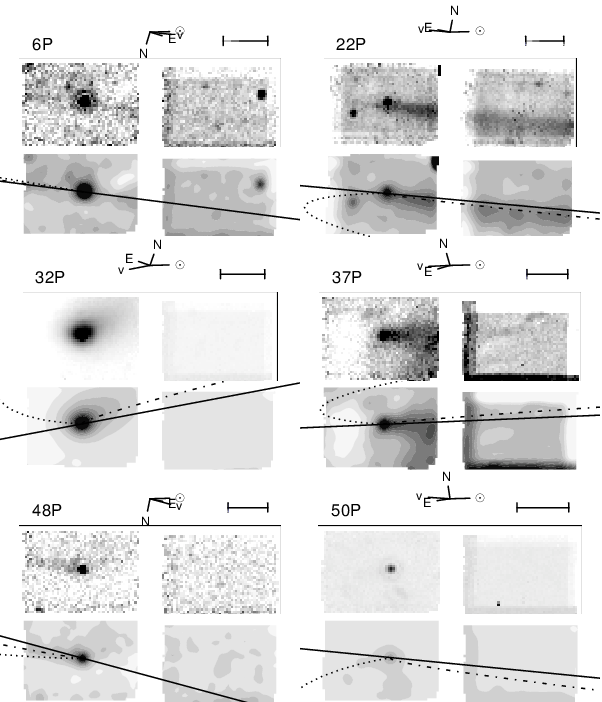}
  \end{center}
  \fi
  \caption{\textit{Spitzer}/IRS 22-\micron{} peak-up images and model
    syndynes for comets with resolved dust (the 16-\micron{} images
    have similar morphologies, but lower signal-to-noise ratios).
    There are two panels for each comet: the lower panels are filled
    contour plots of Gaussian smoothed ($\sigma=1.5$~pixel) versions
    of the upper panels.  Also plotted in the lower panels are
    $\beta=0.1$ (dotted lines) and $\beta=0.001$ (dash-dotted lines)
    syndynes, and the projected orbit of the comet (solid lines).  The
    nuclei are located at the intersections of the three lines.  The
    right images are the 22-\micron{} data obtained while the
    16-\micron{} array was centered on the comet.  In some cases,
    these bonus images contain dust (e.g., 22P).  The top and bottom
    panels are plotted with the same reverse gray-scale data limits.
    The image orientations are indicated: Celestial North (N) and East
    (E), projected sunward direction ($\odot$), and projected velocity
    (v).  Comets 6P/d'Arrest, 22P/Kopff, 32P/Comas Sol\`a, 37P/Forbes,
    48P/Johnson, and 50P/Arend are shown.\label{fig:irs1}}
\end{figure}

\begin{figure}
  \ifincludegraphics
  \begin{center}
    \includegraphics[width=\columnwidth]{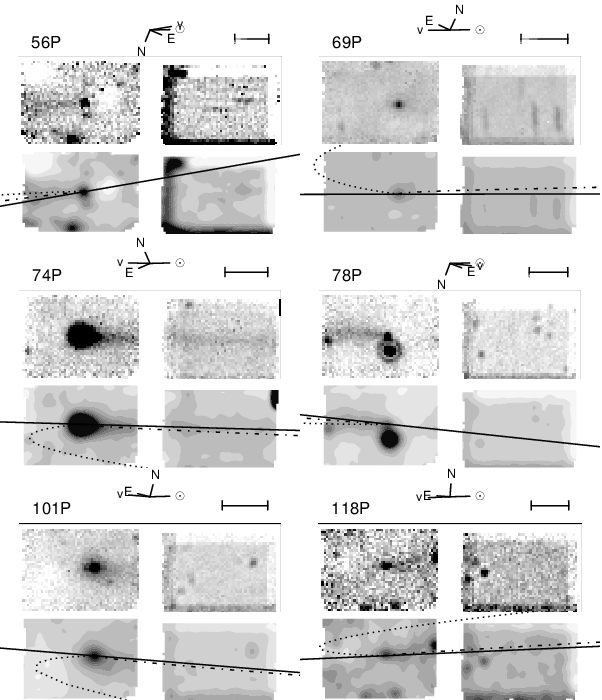}
  \end{center}
  \fi
  \caption{Same as Fig.~\ref{fig:irs1}, but for Comets
    56P/Slaughter-Burnham, 69P/Tay\-lor, 74P/Smir\-no\-va-Cher\-nykh,
    78P/Gehr\-els~2, 101P/Cher\-nykh, and
    118P/Shoe\-ma\-ker-Le\-vy~4.\label{fig:irs2}}
\end{figure}

\begin{figure}
  \ifincludegraphics
  \begin{center}
    \includegraphics[width=\columnwidth]{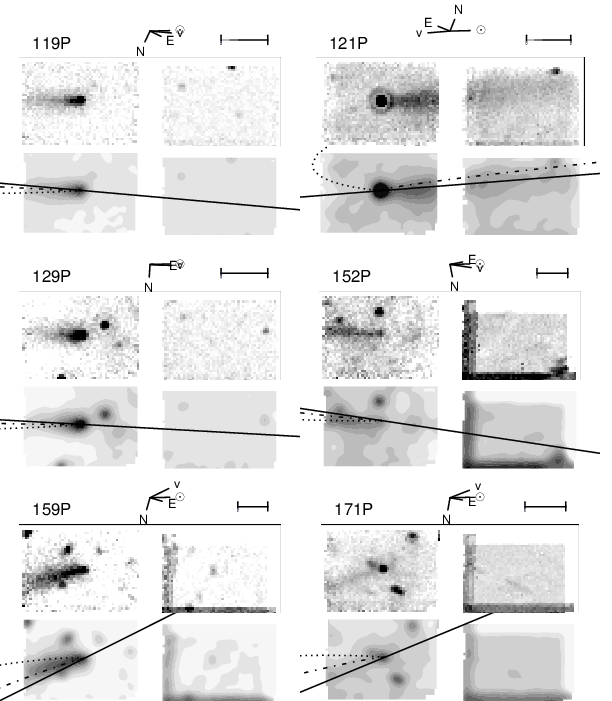}
  \end{center}
  \fi
  \caption{Same as Fig.~\ref{fig:irs1}, but for Comets
    119P/Park\-er-Hart\-ley, 121P/Shoemaker-Holt~2,
    129P/Shoemaker-Levy~3, 152P/Helin-Lawrence, 159P/LONEOS, and
    171P/Spahr.\label{fig:irs3}}
\end{figure}

\begin{figure}
  \ifincludegraphics
  \begin{center}
    \includegraphics[width=\columnwidth]{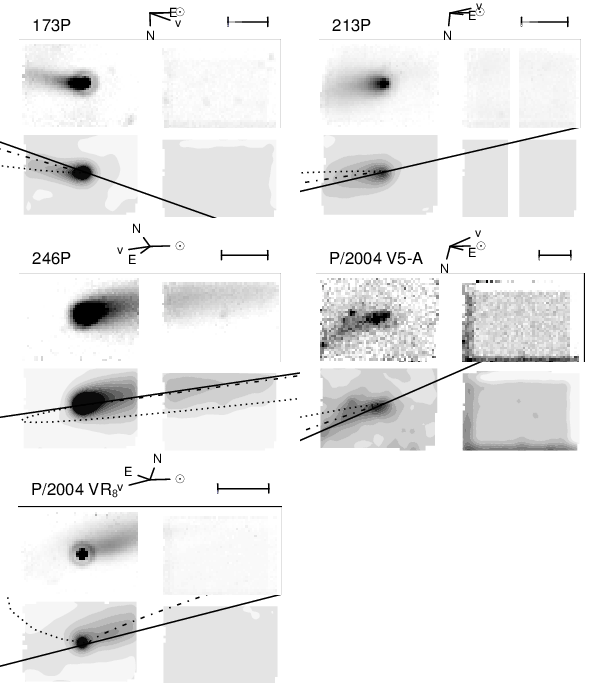}
  \end{center}
  \fi
  \caption{Same as Fig.~\ref{fig:irs1}, but for Comets 173P/Mueller~5,
    213P/2005 R2 (Van Ness), 246P/2004 F3 (NEAT), P/2004 V5-A
    (LINEAR-Hill), and P/2004 VR$_8$ (LONEOS).\label{fig:irs4}}
\end{figure}

\begin{figure}
  \ifincludegraphics
  \begin{center}
    \includegraphics[width=\columnwidth]{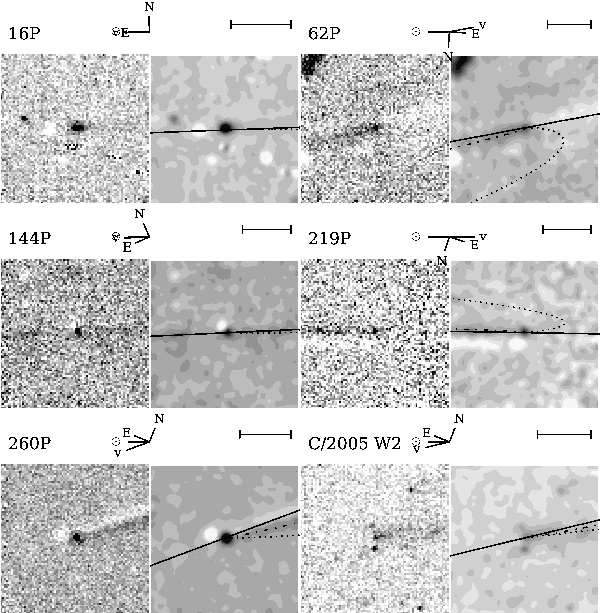}
  \end{center}
  \fi
  \caption{Same as Fig.~\ref{fig:irs1}, but for \textit{Spitzer}/MIPS
    24-\micron{} images.  Most images are background subtracted with
    the second (shadow) observation of the same comet.  In these
    cases, the grayscale data limits are chosen to enhance the
    contrast of only one of the two comet images.  Comets
    16P/Brooks~2, 62P/Tsuchinshan~1, 144P/Kushida, 219P/2002 LZ$_{11}$
    (LINEAR), 260P/2005 K3 (McNaught), and C/2005 W2 (Christensen) are
    shown.\label{fig:mips1}}
\end{figure}

\begin{figure}
  \ifincludegraphics
  \begin{center}
    \includegraphics[width=\columnwidth]{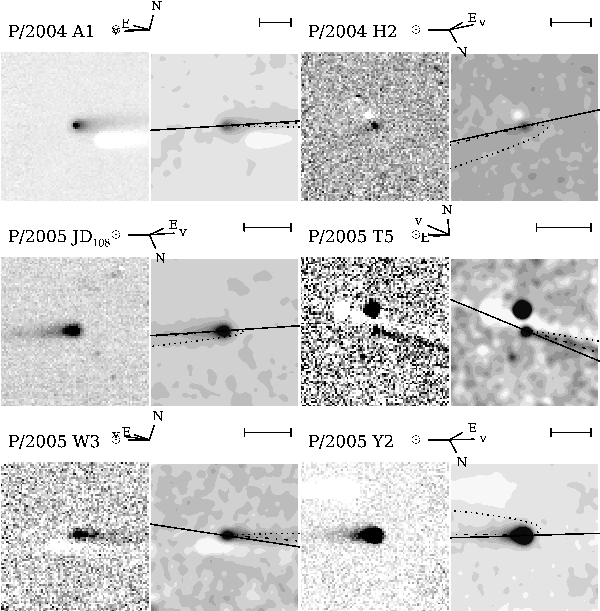}
  \end{center}
  \fi
  \caption{Same as Fig.~\ref{fig:mips1}, but for Comets P/2004 A1
    (LONEOS), P/2004 H2 (Larsen), P/2005 JD$_{108}$ (Catalina-NEAT),
    P/2005 T5 (Broughton), P/2005 W3 (Kowalski), and P/2005 Y2
    (McNaught).\label{fig:mips2}}
\end{figure}

If dust is detected, we describe it as a coma, tail, trail, or some
combination of the three.  To aid our identifications, we generated a
set of zero-ejection-velocity syndynes (curves of constant $\beta$ but
variable emission time), calculated for $\beta=1$, $10^{-1}$,
$10^{-2}$, $10^{-3}$, and $10^{-4}$, using the software of
\citet{kelley06-phd} and \citet{kelley08}.  The survey images with
dust detections are presented in Figs.~\ref{fig:irs1} through
\ref{fig:mips2}, along with model syndynes.  Our identifications are
listed in Table~\ref{tab:phot} and are based on the following
guidelines:
\begin{enumerate}
\item Comets with bright dust surrounding the nucleus are described as
  having comae (e.g., 32P, 74P, 213P).
\item Comets with dust following the $\beta>10^{-3}$ syndynes are
  described as having tails (e.g., 159P, 173P).
\item Comets with long linear dust features following the
  $\beta\leq10^{-3}$ syndynes are described as having trails (e.g.,
  74P, 219P).  We also labeled the dust as a trail if it leads the
  comet along the projected orbit (e.g., 22P, 144P).
\item Comets with thin linear dust features that overlap multiple
  syndynes and the projected orbit of the comet are ambiguous, and are
  described with the label tail/trail (e.g., 56P, 78P).
\item Comets with broad, but short, dust detections along any syndyne
  are labeled as tails (e.g., 119P, P/2005 JD$_{108}$).
\end{enumerate}

Out of 89 comets, 56 comets (63\%) do not have clear morphological
evidence for dust.  The remaining 33 dust detections are described in
Table~\ref{tab:phot}.  In addition, 2 comets are marked as tentative
detections of dust: Comet 6P/d'Arrest may have a very faint trail
along the projected orbit of this comet, and Comet 50P/Arend has a
slight surface brightness excess along the $\beta=0.1$ syndyne (but
there is also a nearby star, which complicates the interpretation).
Both comets are shown in Fig.~\ref{fig:irs1}.  These cases are
``tentative'' as opposed to ``ambiguous'' because dust has not been
definitively detected.

\subsection{Dust photometry}\label{sec:photometry}
The \spitzer{} images allow us to estimate each comet's dust mass and,
for the IRS observations, color temperature.  For comets with a coma
and/or tail, we measure the total thermal emission in 3--6~pixel
radius apertures centered on the comet.  The smaller aperture sizes
are necessary in some images to reduce background contamination.  When
possible, we also examine an aperture offset at least 5~pixels from
the nucleus in order to measure emission from tails and trails.  The
size and shape of the tail/trail aperture varies comet-by-comet and
depends on the morphology of the dust.  The background is estimated
from areas of low contamination from dust and background sources.  As
the thermal emission from comets typically contains significant
emission from the nucleus, we only use the nucleus subtracted images
from \citetalias{fernandez11}.  The fluxes have been aperture
corrected and color corrected assuming the spectral shape of an
isothermal blackbody sphere in local thermodynamic equilibrium (LTE),
according to the methods prescribed by the MIPS and IRS instrument
handbooks \citep{mips, irs}.  The dust fluxes are listed in
Table~\ref{tab:phot}.

With the two IRS peak-up arrays, we measure the 16--22-\micron{} color
temperature of the dust in our survey, centered on the nucleus and in
the tail/trail (if applicable) for all fluxes with signal-to-noise
ratios $\geq5$.  Dust color temperatures are listed in
Table~\ref{tab:color} as a ratio with the temperature of an isothermal
blackbody sphere in LTE at the same heliocentric distance ($T_{BB} =
278\,r_h^{-1/2}$).  The error-weighted mean of all color-temperature
measurements is $\langle T/T_{BB} \rangle = 1.074 \pm 0.006$, of the
center apertures is $1.079 \pm 0.007$, and of the tail/trail apertures
is $1.065 \pm 0.010$.

The color temperature of the dust will only serve as an approximation
to the true physical temperature of the grains because physical
temperatures depend on the grain sizes, shapes, and compositions, and
comet dust structures are collections of grains with a wide range of
properties.  For example, dust grains with sizes $\lesssim
1$~\micron{} have higher equilibrium temperatures than larger grains
because the former have smaller emission cross sections at
10--20~\micron, the wavelength regime at which the peak of the thermal
emission occurs for a blackbody sphere in the inner solar system.
Thus, their equilibrium temperatures are increased over that of a
blackbody.  These increases are composition dependent as carbonaceous
grains, which have high absorption and emission efficiencies, will
have different equilibrium temperatures than an equally-sized silicate
grain, which is less absorptive at most optical and infrared
wavelengths.  For more discussion on these effects, and how they may
constrain thermal emission spectra, see \citet{wooden02}.

Despite the ambiguity in correlating color temperatures to coma grain
properties, comet-to-comet differences indicate variations of those
(unknown) dust parameters among the members of our survey.  For
example, the color temperature of 74P/Smirnova-Chernykh's coma is
greater than 101P-A/Chernykh's coma ($1.16\pm0.02$ versus
$0.96\pm0.03$, respectively).  This difference indicates that 74P's
coma is composed of different grains than 101P's coma, and we can
speculate that they are smaller and/or more carbon rich in the former.

\subsection{Survey sensitivity}\label{sec:sens}
\subsubsection{All dust}
The SEPPCoN survey was designed to detect comet nuclei with a
signal-to-noise ratio of 30.  The exposure times are not based on
potential dust emission but solely on our estimated nucleus sizes and
temperatures.  To verify that we can draw meaningful results on the
dust activity of Jupiter-family comets as a whole, we must estimate
the survey's sensitivity to dust.  Rather than using a theoretical
estimate (i.e., observation planning tools), we prefer to measure the
sensitivities from our observations, which will easily take into
account the effects of background objects (stars, asteroids, etc.),
variances in instrument calibration (e.g., flat-fielding), latent
charge on the detector, etc.  To measure the sensitivity of an image,
we first filter the image to remove structures larger than
$5\times5$~pixels (i.e., dust and the smoothly varying background).
To be more specific, we applied a morphological gray closing operator,
followed by the opening operator, and subtracted the result from the
original image.  The opening and closing operators effectively
convolve the image with a $5\times5$ box, but instead of replacing
each pixel with the average of the surrounding pixels, the closing and
opening operators replace each pixel with the maximum and minimum
pixel values within the box, respectively.  Applying the closing and
opening operators is similar to applying a median filter, except more
of the small scale structure is lost in median filtering.  Because
small scale structure affects our ability to detect dust, we prefer
the closing and opening filters over the median.  See \citet{lea89}
and \citet{appleton93} for further discussion and other applications
of morphological operators in astronomy.  In the part of the co-added
image where the integration time was the highest, we measure the
standard deviation of the image after iteratively removing $3\sigma$
outliers (i.e., the comet itself, and the central cores of stars).

The image sensitivities are measured in units of \mjysr{}.  However,
dust surface brightness depends on heliocentric distance: a surface
brightness of 1~\mjysr{} at 3~AU implies less dust than 1~\mjysr{} at
7~AU due to the different equilibrium temperatures.  A more relevant
representation of the sensitivity is needed.  We have chosen to
transform the measured surface brightnesses into optical depths.  To
compute the image sensitivity to dust in terms of optical depth, we
use the equation:
\begin{equation}
  \sigma_\tau = \frac{\sigma_{I_\nu}}{C(T_d) B_\nu(T_d) \sqrt{A}},
\end{equation}
where $\sigma_\tau$ is the optical depth (or effective fill factor) of
a $1\sigma$ per pixel detection of dust, $\sigma_{I_\nu}$ is the
measured sensitivity ($1\sigma$ per pixel) in units of \mjysr, $T_d$
is the effective temperature of the dust, $C(T_d)$ is the instrument
color correction for a blackbody spectrum at a temperature $T_d$,
$B_\nu(T_d)$ is the Planck function evaluated at temperature $T_d$ in
units of \mjysr, and $A$ is the area of a fictitious dust aperture in
units of square pixels.  Our default aperture size in
Table~\ref{tab:phot} is a 6~pixel radius circle.  In
\S\ref{sec:photometry}, we computed a mean color temperature of
$T_c/T_{BB} =1.08$ in the center aperture.  A priori, we do not know
the color temperature of any specific comet.  Rather than using one
value for 70 comets and our measured values for those 10 comets in
Table~\ref{tab:color}, we assume $T_d \approx T_c = 1.08\,T_{BB} =
300\,r_h^{-1/2}$~K for all comets.

Histograms of the observed dust optical depths and the survey image
sensitivities are presented in Fig.~\ref{fig:sens} for the IRS
22-\micron{} and MIPS 24-\micron{} observations.  Notice that the dust
detections fall off at the same optical depths as our estimated
$3\sigma$ image sensitivities.  This fall off strongly suggests our
dust detections are sensitivity limited.

\subsubsection{Trail dust}
In a survey of 34 comets taken with the MIPS instrument,
\citet{reach07} found that at least 80\% of all short-period comets
have dust trails.  Since dust trails are long lived features
(\S\ref{sec:morphology}) we may expect a similar rate for the SEPPCoN
targets, but instead we find a rate of 10\%.

To better understand the dramatic difference in trail rates, we
compared our survey target list to that in \citeauthor{reach07} and
found the following 12 comets in both surveys.
\begin{itemize}
\item Comet 107P did not have a trail, or any other dust, in any
  observation.

\item Comet 62P had a trail in both surveys.

\item Comet 78P's dust was classified as a possible trail in SEPPCoN,
  and as having a trail by \citeauthor{reach07}.

\item Comet 121P had a trail in SEPPCoN but only a tail in the
  \citeauthor{reach07} survey.  We suggest that this comet's trail
  only becomes apparent at large true anomalies (trail survey
  $f=31\degr$, SEPPCoN $f=123\degr$).

\item Comets 32P, 69P, 123P, and 131P had trails in the
  \citeauthor{reach07} survey, but they appear to be too faint to detect in
  the SEPPCoN images.  In particular, 123P's SEPPCoN images have
  several nearby point sources and background subtraction artifacts
  \citep{fernandez11}.

\item Comets 48P and 129P had clear leading and following trails in
  the \citeauthor{reach07} survey.  The fields-of-view of the SEPPCoN
  observations are limited in the trailing direction and tails overlap
  with the orbit.  It is surprising that the leading trails are not
  observed in either case.  The leading trail may be a transient
  feature.

\item Comets 94P and 127P were classified as an ``intermediate trail''
  by \citeauthor{reach07}, i.e., the dust more closely followed the
  $\beta=10^{-3}$ syndyne than the projected orbit of the comet (this
  label is consistent with our ``trail'' label).  Neither of these
  comets have dust in our survey images and both were taken at larger
  true anomalies, emphasizing that such dust may be relatively short
  lived (trail survey $f=62\degr$ and 77\degr, SEPPCoN $f=150\degr$
  and -152\degr).

\end{itemize}
In summary, out of the 12 comets in both surveys, three morphologies
have consistent descriptions (62P, 78P, and 107P), four trails appear
to be too faint for SEPPCoN (32P, 69P, 123P, and 131P), two
intermediate trails appear to be short lived and limited to smaller
true anomalies (94P and 127P), one comet's trail may not be present at
very small true anomalies (121P), and two trails may be obfuscated by
dust tails, whereas their leading trails are possibly transient
features (48P and 129P).  From this comet-by-comet comparison, it
appears the low trail detection rate of SEPPCoN can be explained by:
1) observation timing/geometry; 2) the limited field-of-view of the
IRS peak-up arrays; and, 3) the survey's sensitivity to dust.  The
last point is discussed further below.

\begin{figure}
  \ifincludegraphics
  \begin{center}
    \includegraphics[width=\columnwidth]{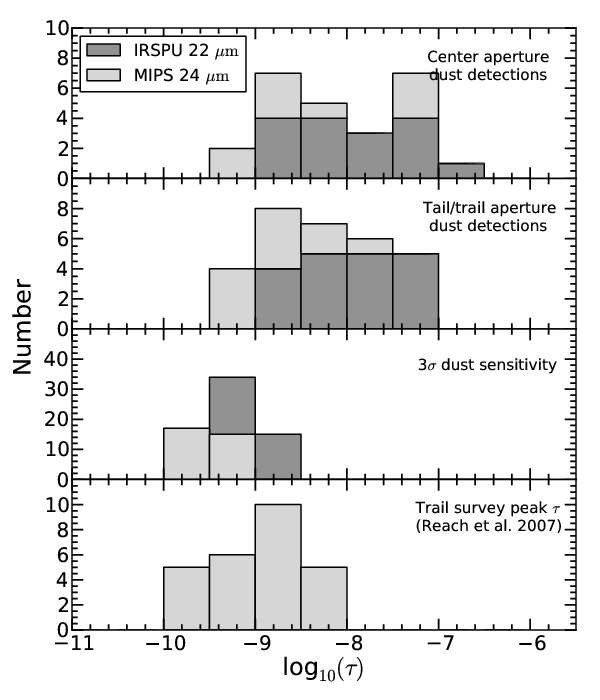}
  \end{center}
  \fi
  \caption{Histograms of the IRS 22-\micron{} and MIPS 24-\micron{}
    center (top panel) and tail/trail (upper-middle panel) photometry
    expressed as dust optical depth.  Also shown are dust
    sensitivities ($3\sigma$ per 6-pixel radius aperture) in terms of
    optical depth for all IRS 22-\micron{} and MIPS 24-\micron{}
    images (lower-middle panel) and \spitzer{} trail survey peak
    optical depths (bottom panel) from Table~2 of
    \citet{reach07}.\label{fig:sens}}
\end{figure}

A portion of the difference in trail detection rates is due to the
typical observing geometries of short-period comets at
$r_h\gtrsim4$~AU, which cause tails and trails to overlap on the sky
(e.g., compare 22P to 48P in Fig.~\ref{fig:irs1}).  However,
obfuscation from brighter coma and tail dust does not explain the lack
of trails in the 56 images of apparently bare nuclei.  Instead, we
must consider the survey differences in sensitivity.  First note that
the \citeauthor{reach07} survey targeted comets within 3.5~AU from the
Sun, whereas the SEPPCoN observations were all at $r_h\geq3.4$~AU.
Also note that the trail survey integration times were shallow (either
42 or 140~s on source with MIPS) compared to the SEPPCoN observations
(140 to 2500~s with MIPS).  We can make direct comparisons between the
surveys by converting trail fluxes into optical depths.  We examined
the peak $\tau$ values listed in Table~2 of \citeauthor{reach07}, and
find the trails have $0.2\ten{-9} \leq \tau \leq 9\ten{-9}$ in the
mid-IR, with a median of 1.2\ten{-9}.  For the eight SEPPCoN targets
with trail detections, we find $0.4\ten{-9} \leq \tau \leq
11\ten{-9}$, and a median of 1.3\ten{-9}.  In Fig.~\ref{fig:sens}, we
also show a histogram of the peak optical depths from the
\citeauthor{reach07} survey.  The range and median trail optical
depths of the two surveys roughly agree.  But the \citeauthor{reach07}
trail survey did detect most of the faintest trails, and this is
likely the main difference between the two surveys.

\section{Discussion}\label{sec:discussion}
\subsection{Comet activity}\label{sec:activity}
Comet activity is controlled by physical and dynamical processes at
the nucleus.  Physical properties, such as composition and geology,
determine if a comet is active under a given set of solar illumination
conditions.  Besides size and color temperature, little is known about
the physical properties of our targets.  The dynamical properties of a
nucleus are described by its orbit and rotation state, and together
they also determine the solar illumination conditions and history.
The rotation states are known for only a few of our targets.  In
contrast, all of the orbital parameters are constrained well enough to
make meaningful comparisons.  The geometrical circumstances of each
observation are also well constrained and can be tested.  For example,
we might test for a correlation between activity and observer-comet
distance because dust around comets farther from the telescope is more
difficult to detect (lower spatial resolution, possibly cooler dust
temperatures).

For this work we have tested for the presence of a correlation between
recent activity and eight observational and orbital parameters: $r_h$,
$\Delta_S$ (\spitzer-comet distance), $\phi_S$ (phase angle,
Sun-comet-\spitzer), $f$ (true anomaly), $q$ (perihelion distance),
$e$ (orbital eccentricity), $a$ (orbital semi-major axis), and $\Delta
q|_{150}$ (perihelion distance history, described below).  All orbital
parameters are measured from their osculating elements at the time of
the observation, except $\Delta q|_{150}$.  We integrated each comet's
current orbit back 300~yr in 90~day steps using HORIZONS
\citep{giorgini96}; $\Delta q|_{150}$ is the difference between the
minimum and maximum $q$ over the past 150~yr,
\begin{equation}
  \Delta q = \mbox{min}(q) - \mbox{max}(q)
\end{equation}
where a negative $\Delta q|_{150}$ indicates the perihelion distance
was larger in the past.  A relationship between any one of the above
parameters and activity is not necessarily expected (e.g., we don't
expect activity to depend on the phase angle of the observation), but
we tested for correlations in order to perform an unbiased study while
getting a useful sense of what a ``null hypothesis'' result looks
like.

For each parameter, we compute a two-sided Kolmogorov-Smirnov (K-S)
statistic, $D$, and two-tailed $p$-value with the null hypothesis that
the active and inactive comets in our survey are drawn from the same
distribution.  The $p$-value (or false positive probability) indicates
the probability that uncorrelated data sets might result yield a $D$
greater than that observed.  When $D$ is small, or when the $p$-value
is large, the null hypothesis cannot be rejected.  At first, only
those comets with comae and/or tails were considered to be active.
Comets with trails, or without dust are considered inactive.  The four
comets with ambiguous morphologies (tail/trail) are dropped from the
analysis.  With these definitions, the numbers of active and inactive
comets are 21 and 64.  Our K-S test results are listed in
Table~\ref{tab:kstests}.

\begin{figure}
  \ifincludegraphics
  \begin{center}
    \includegraphics[width=\columnwidth]{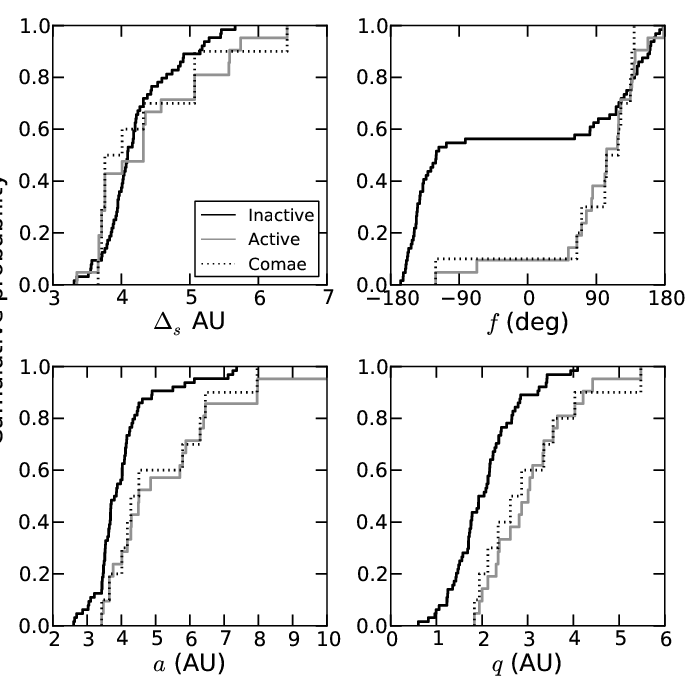}
  \end{center}
  \fi
  \caption{Cumulative probability functions comparing the
    distributions of \spitzer-comet distance ($\Delta_S$), true
    anomaly ($f$), orbital semi-major axis ($a$), and perihelion
    distance ($q$) for active (coma and/or tail) and inactive comets
    in our survey.  The active distributions based solely on the
    presence of a coma are also shown.\label{fig:kstests}}
\end{figure}

\begin{figure}
  \ifincludegraphics
  \centering

  \includegraphics[width=0.47\columnwidth]{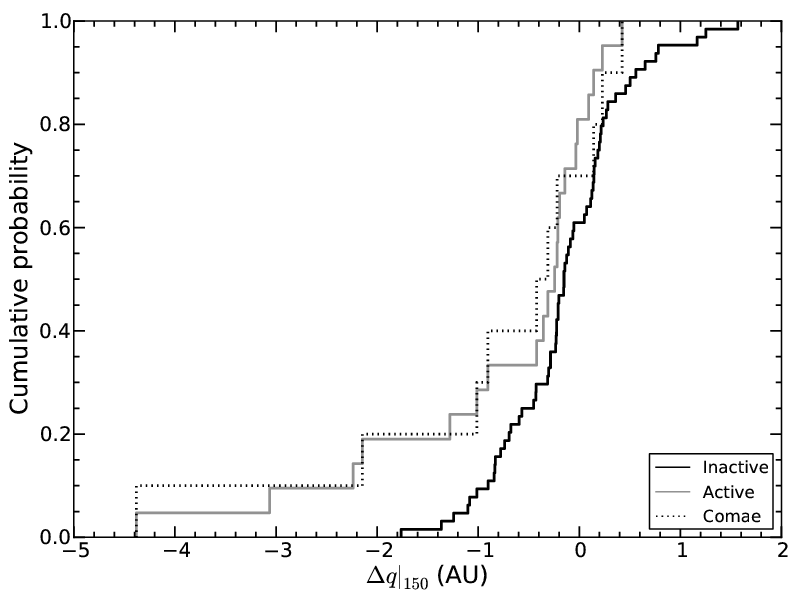}
  \hfill
  \includegraphics[width=0.47\columnwidth]{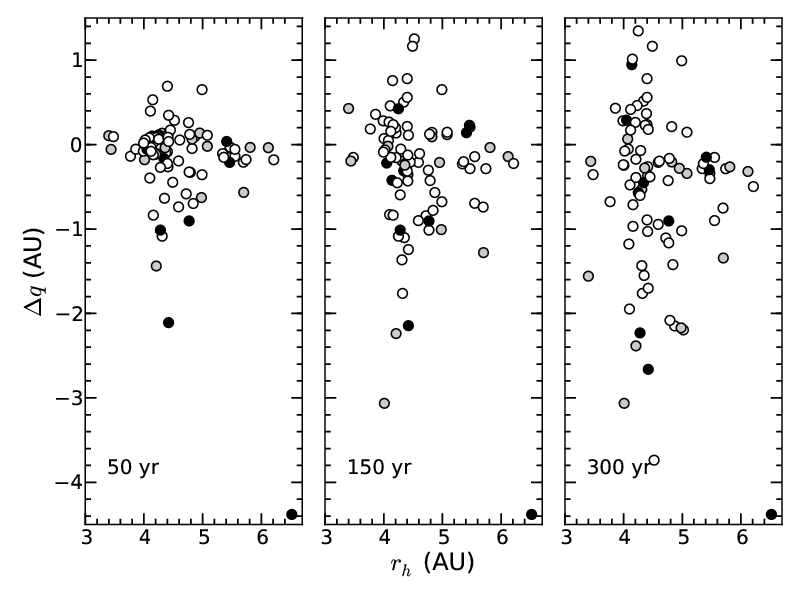}

  \vspace{1em}
  \includegraphics[width=0.47\columnwidth]{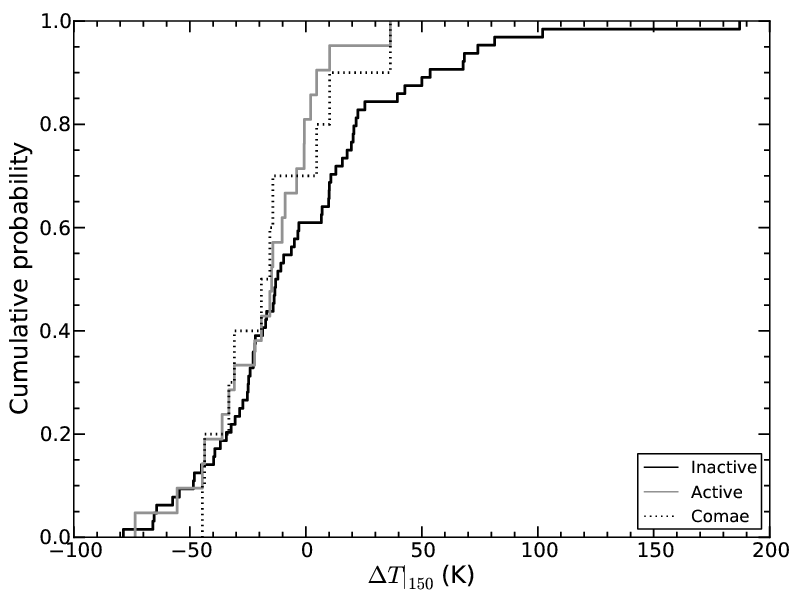}
  \hfill
  \includegraphics[width=0.47\columnwidth]{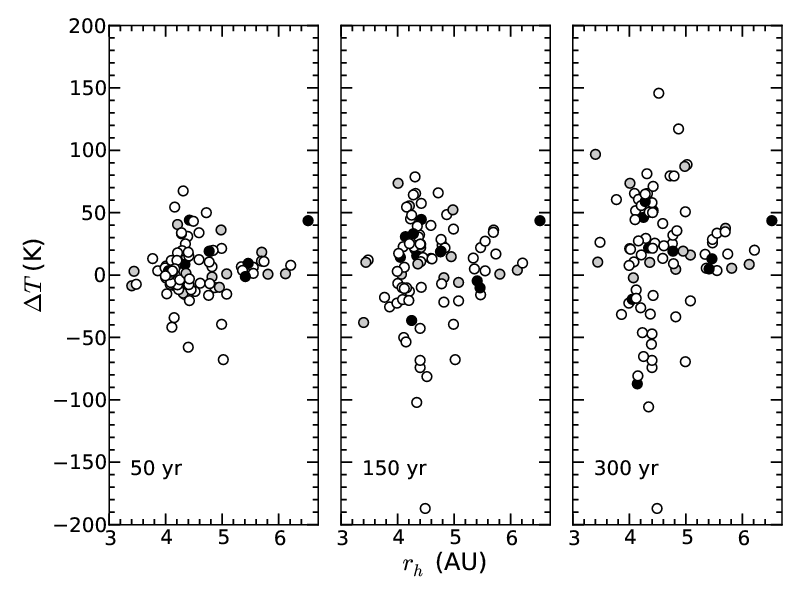}
  \fi

  \caption{(Top left) Cumulative probability functions comparing the
    distributions of perihelion distance variations in the past 150~yr
    ($\Delta q|_{150}$) for active and inactive comets in our survey.
    (Top right) Plot of $\Delta q$ versus observed $r_h$ computed for
    50-, 150-, and 300-yr orbital histories (comae: black, tails:
    gray, inactive: white).  (Bottom left) Cumulative probability
    functions for peak-sub-solar-temperature variations in the past
    150~yr ($\Delta T|_{150}$).  (Bottom right) Plot of $\Delta T$
    versus observed $r_h$ computed for 50-, 150-, and 300-yr orbital
    histories. \label{fig:deltaq}}
\end{figure}

In \S\ref{sec:morphology}, we argue that dust tails are associated
with activity on longer timescales than dust comae, and that they do
not necessarily indicate a currently active comet.  As an additional
check on the significance of our K-S tests, we removed tail-only
detections from our active comet list, and compared the remaining
comets to the inactive comets.  This change reduced the number of
active comets to 10.  The second set of K-S tests are also presented
in Table~\ref{tab:kstests}.

The $p$-values in Table~\ref{tab:kstests} indicate that the active and
inactive comets have significantly different $f$, $q$, $e$, and $a$
distributions.  The distribution of $\Delta_S$ is slightly different
but with less confidence, indicating that comet-\spitzer{} distance
does not have a strong influence on our results.
Figure~\ref{fig:kstests} presents the cumulative probability functions
for $f$, $q$, $a$, and $\Delta_S$.  The cumulative probability
functions for $\Delta q|_{150}$ are presented in
Fig.~\ref{fig:deltaq}.  We discuss each result below.

\subsubsection{True anomaly}\label{sec:f}
The active and inactive comets have significantly different
distributions in true anomaly.  In Fig.~\ref{fig:kstests}, we see that
the active comets are strongly biased toward positive true anomalies
(i.e., post-perihelion), with only 2 active comets at $f<0$\degr{}
(i.e., pre-perihelion).  Restricting our definition of active to only
those comets with a coma does not diminish our conclusion.  In
addition, no comet in our survey is active between $f=-180$\degr{} and
$-125$\degr.

\begin{figure}
  \ifincludegraphics
  \includegraphics[width=\columnwidth]{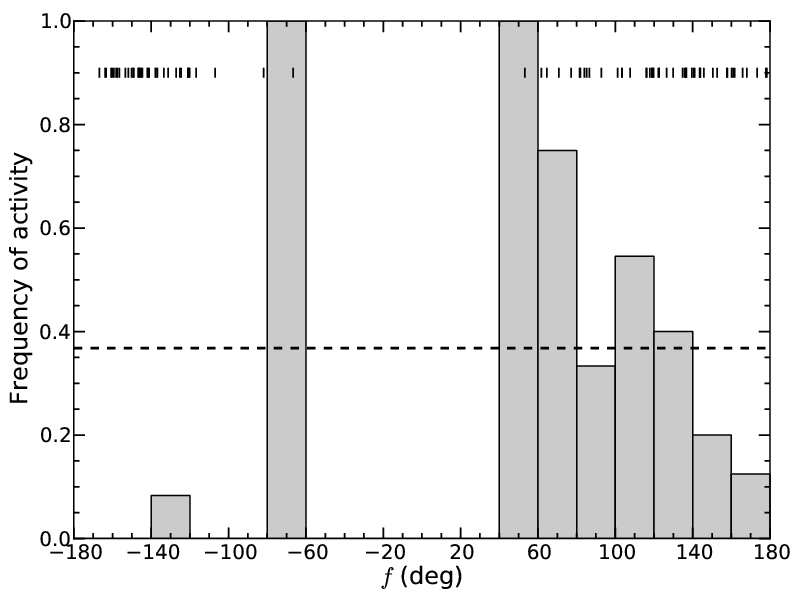}
  \fi

  \caption{Frequency of activity versus true anomaly.  A horizontal
    line is plotted at $e^{-1}$ in order to estimate at which true
    anomalies activity is typically initiated and quenched.  Short
    vertical lines mark the true anomalies of individual
    targets.\label{fig:tahist}}
\end{figure}

As an alternative metric of post-perihelion activity, we estimate the
true anomalies at which activity starts and stops.
Figure~\ref{fig:tahist} shows a histogram of activity versus true
anomaly, with a horizontal line plotted at $\exp(-1)$.  We chose
$\exp(-1)$ because it marks the half length of an exponential
fall-off, however we note that our histogram does not necessarily
suggest this particular functional relationship between activity and
true anomaly.  By comparing the line to the histogram, we estimate
that activity in the average JFC turns on at $f>-120\degr$ and turns
off at $f\approx140\degr$.

A few scenarios may result in apparent persistent post-perihelion
activity: 1) a temporary dust mantle insulates the ices during the
pre-perihelion portion of the orbit but is lost near perihelion
allowing persistent activity on the outbound leg of the orbit; 2) the
surface layers retain a significant seasonal thermal wave that
continues to heat the comet sub-surface when insolation would normally
be insufficient; 3) a seasonal effect near perihelion that causes
persistent activity well outside of perihelion; or, 4) asymmetric
activity caused by the late onset of volatile production.  For the
purposes of this paper, we define a dust mantle to be an ice-free
layer at the surface of the nucleus.

As solar insolation warms the surface of a comet a thermal wave
propagates inward.  At some depth the thermal wave heats ices to
sublimation temperatures, which drives mass loss.  The depth of those
ices, the thermal properties of the layers above them, and the
rotational and orbital states of the nucleus will determine when
activity is initiated.  Suppose that activity becomes so vigorous that
any putative dust mantle is ejected, either partially or completely,
from the comet.  As the comet recedes from the Sun, activity would be
stronger than the pre-perihelion orbit, but it would eventually
decrease to the point at which it can no longer lift some grains from
the surface, perhaps because of their large sizes, and a dust mantle
will redevelop.  Thus, a pre-perihelion nucleus could be insulated by
a dust mantle, whereas the post-perihelion nucleus might not.
\citet{brin79} present one of the first 1-D models of comet mantle
formation and destruction that causes pre- and post-perihelion
asymmetries near $r_h=1$~AU.  Mantle formation and destruction in
nucleus thermal models is reviewed by \citet{prialnik04}.

In the second scenario, where activity is driven by seasonal heating,
the dust mantle is not lost but retains heat from the perihelion
passage.  Energy deposited on diurnal timescales will radiate back to
space on roughly the same timescale, but sub-surface temperatures can
build up on longer timescales as a comet approaches perihelion.  After
the comet passes perihelion, the thermal wave cools to both space and
the comet sub-surface.  If sub-surface ices continue to be warmed to
sublimation temperatures, despite the increasing heliocentric
distance, activity will persist.  In order to observe a
pre-/post-perihelion asymmetry, the ices must be buried at depths
closer to the annual thermal skin depth than the diurnal thermal skin
depth.  If this is true, gas production should not vary diurnally at
the intermediate heliocentric distances in our survey, or, at least,
it should be weakly correlated with rotational insolation.  There is
limited support for this scenario in flyby images of nuclei.  Jet
activity occurs from surfaces in local night at 81P/Wild~2,
9P/Tempel~1, and 103P/Hartley~2 \citep{sekanina04-w2, farnham07,
  ahearn11}, but the apparent timescales of this activity are much
shorter than what is required for a pre-/post-perihelion asymmetry.

The near-surface structure of comets was directly examined by the
\di{} mission \citep{ahearn05}.  \di{} excavated a crater with a
diameter of 50--200~m on Comet 9P/Tempel~1 \citep{richardson13,
  schultz13}.  According to \citet{ahearn08}, most observations of the
ejecta are consistent with dust-to-volatile ratios of order unity, and
\citet{groussin10} came to the same conclusion.  Since the
dust-to-volatile ratio is not $\gg1$, it appears the thermal wave in
this nucleus had not sublimated much of the water ice on the scale of
the crater depth.  Even greater constraints on the depth of thermal
penetration were obtained with \di{} spectra of the comet's surface.
\citet{groussin07} analyzed the thermal emission from the surface and
showed that it has a low thermal inertia, nearly in instantaneous
equilibrium with sunlight.  They argue that cometary activity must
originate within the first centimeters to meters of the surface.  From
the thermal inertia upper-limits and assumptions on conductivity and
heat capacity, \citet{ahearn08} computes the diurnal skin depth to be
3~cm, and the annual skin depth to be 90~cm.  These numbers are order
of magnitude consistent with theoretical models of comet nuclei made
before the \di{} mission results \citep{prialnik04}, and with
\spitzer{} mid-IR spectral measurements of the nucleus thermal
emission at 5~AU \citep{lisse05}.  In order for dust mantles to quench
activity on diurnal timescales, they must be larger than $\sim1$~cm
thick, and to prevent the seasonal thermal wave from driving activity,
volatiles must be more than $\sim10$~cm from the surface.

\citet{weissman87} proposed that the sudden illumination of a nucleus
hemisphere caused by the rapid change of season near perihelion causes
cracks in the mantle, and these cracks create new active areas on the
nucleus.  The hemisphere illuminated on the in-bound leg of the orbit
does not develop these cracks because it is gradually heated, and the
surface can better accommodate the changes.  If this hemisphere
continues to be illuminated as the comet recedes from the Sun, the
comet's activity could be greater than that of it's inbound leg.

\citet{meech11-epoxi} propose that the pre-/post-perihelion asymmetry
of Comet 103P/Hartley 2 is caused by the late on-set of \coo{} driven
activity.  They find that this comet's activity was first initiated in
2010 at 4.3~AU by water sublimation.  When the comet reached 1.4~AU
(pre-perihelion) \coo{} driven activity dominated, to the point at
which water ice can be driven off the surface, as if it were dust.
The \coo{} driven activity would persist beyond the comet's aphelion
\citep{meech11-epoxi}.

Whatever the cause, the effect is clear in the SEPPCoN.  Because the
survey took a consistent approach to its observations, it can be used
to test these and other inhibitors and drivers of activity in a
statistical manner.

\subsubsection{Perihelion distance}\label{sec:q}
The inactive comets have smaller perihelion distances than the active
comets.  The median $q$ is 2.0~AU for the inactive comets and 3.0~AU
for the active comets.  This difference may be a selection effect.
Cometary activity is driven by insolation, and, in general, comets are
most likely to be active when they are near perihelion.  Our survey
targets were observed at heliocentric distances ranging from 3 to
7~AU.  However, a comet with a perihelion distance between 3 and 7~AU
is \textit{a priori} more likely to be active in our survey, since its
activity at such distances has already been established (otherwise it
would not have been designated a comet, and not observed by SEPPCoN).
Therefore, the greater rate of activity for comets with larger
perihelion distances appears to partially be an artifact of the
survey.

\begin{figure}
  \ifincludegraphics
  \begin{center}
    \includegraphics[width=\columnwidth]{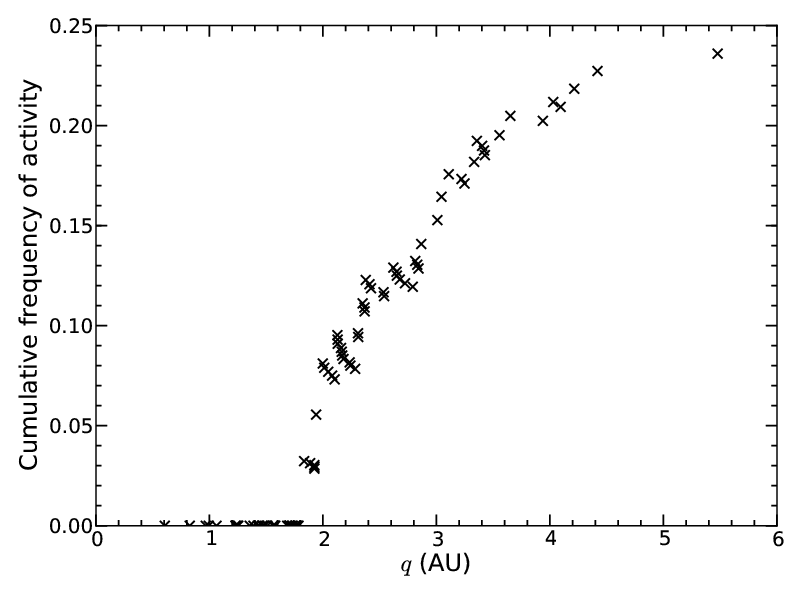}
  \end{center}
  \fi
  \caption{The cumulative frequency of activity in our survey versus
    perihelion distance.\label{fig:cumulative-activity}}
\end{figure}

To further demonstrate the larger activity rates of the large
perihelia comets, we plot the cumulative frequency of activity in our
survey, sorted by perihelion distance, in
Fig.~\ref{fig:cumulative-activity}.  That the frequency of activity
almost continuously increases for $q>2$~AU shows the higher activity
rate of this population.

\begin{figure}
  \ifincludegraphics
  \begin{center}
    \includegraphics[width=\columnwidth]{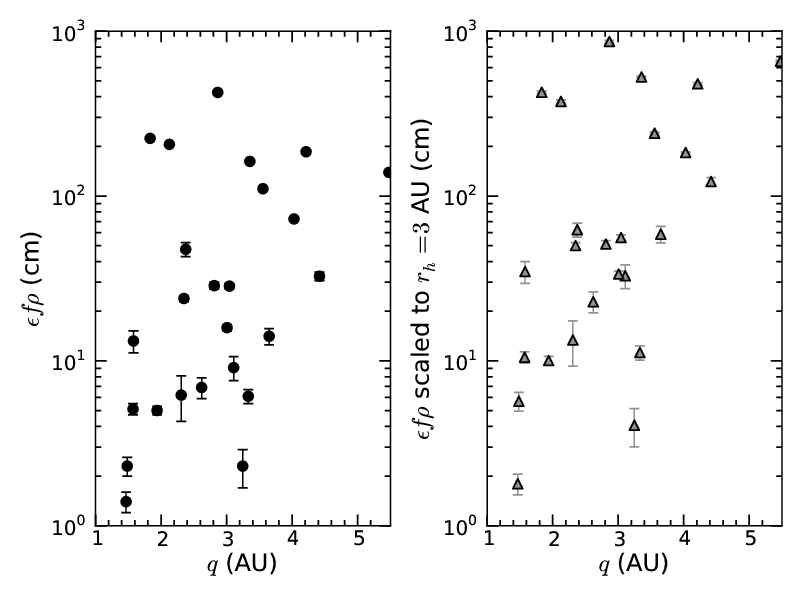}
  \end{center}
  \fi
  \caption{(Left) Comet activity, quantified with the parameter
    \efrho, versus perihelion distance.  (Right) The same as on the
    left, but \efrho{} has been scaled by $(r_h / 3~{\rm AU})^2$, so
    that comets can be more directly compared to each other.  Note the
    paucity of low-\efrho{} comets with large perihelion distances,
    indicating that our survey and the discovered JFC population is
    sensitivity limited. \label{fig:efrho-v-q}}
\end{figure}

Figure~\ref{fig:efrho-v-q} presents an alternative view on these large
perihelia comets, and shows that on average they produce more dust
than the smaller perihelia comets.  Here, we plot \efrho{} versus
perihelion distance.  We define the parameter \efrho{} as the product
of effective emissivity ($\epsilon$), filling factor ($f$), and
projected aperture radius ($\rho$, in units of cm).  This parameter is
intended to be analogous to the \afrho{} parameter commonly computed
for scattered light observations \citep{ahearn84}.  Further discussion
of \efrho{} can be found in \ref{sec:efrho}.  We based our \efrho{}
values on the mean MIPS 24-\micron{}, or the IRS 22-\micron{} center
aperture photometry (Table~\ref{tab:efrho}).  Comets with small
\efrho{} values, i.e., low-activity comets, are missing from the
large-perihelion population.  We suggest that this bias is due to the
discovery circumstances of large-perihelia comets, and that
low-activity comets with large-perihelia are as yet missing from the
discovered JFC population.

Although we cannot make definite conclusions on the activity of comets
with large-perihelia, we can consider the activity of small-perihelia
comets, whose discovery rates are likely more complete and less
biased.  Of those comets with $q<2.5$~AU, 7 out of 59 comets (12\%)
are active, and none of the 30 comets with $q<1.8$~AU are active
(Fig.~\ref{fig:cumulative-activity}).\footnote{Because the strong bias
  in activity to post-perihelion epochs, we verified that the 30
  comets with $q<1.8$~AU have the same true anomaly distribution as
  comets with $q>1.8$~AU.  A K-S test yields a 1\% probability that
  the two populations are different.}  Therefore any short-period
comet with a small perihelion distance is likely inactive at 3--7~AU.
This finding may be explained if relatively more volatile ices are
lost for comets that approach the Sun more closely during their
perihelion passages, but this hypothesis seems inconsistent with the
results of \citet{ahearn12-origins}.  They studied the \water, \coo,
and CO content of comets in the literature, and found no correlation
with orbital parameters, except for a decreasing \coo{}-to-\water{}
mixing ratio for comets with decreasing perihelion distance, for
$q<1$~AU.  They rightly dismiss this correlation as due to small
number statistics (only 4 comets have a measured \coo{} abundance for
$q<1$~AU) and for the fact that it contradicts their result that
CO/\water{} is uncorrelated with $q$.  Taking our results together
with the discussion of \citet{ahearn12-origins}, we suggest that
comets with small perihelion distances have a thicker insulating
surface than comets with larger perihelion distances.

\subsubsection{Perihelion distance history}\label{sec:deltaq}
\citet{licandro00} surveyed 18 comets in order to estimate the
effective sizes of their nuclei.  Seven of their targets were active:
one at $r_h=3.04$~AU, and the rest at $r_h>4$~AU.  They examined their
orbital histories and found that all but one of their comets recently
perturbed to orbits with smaller perihelion distances in the past
150~yr ($\Delta q|_{150}<-1$~AU) were active.  \citeauthor{licandro00}
hypothesize that the perturbations to smaller perihelion distances
caused a marginally stable mantle to be destroyed, exposing fresh
ices.  The absence of a mantle allows activity to persist to larger
than usual perihelion distances.  When we examined the
pre-/post-perihelion asymmetry in SEPPCoN, we also introduced the
concept of a temporary mantle, only in our case the mantle is
partially or completely lost each perihelion passage.  We now examine
the orbital histories of SEPPCoN targets to further investigate the
potential existence of dust mantles.  Note that a small $\Delta q$ is
not the same as small $q$, i.e., a comet may be perturbed from the
Centaur region ($a > 5.2$~AU) into the Jupiter-family comet region ($a
< 5.2$~AU), yielding $\Delta q < 0$, yet still have a large perihelion
distance in our survey.

Recognizing that the same $\Delta q$ is more significant at small
perihelion distances than at large perihelion distances, we have also
computed the maximum sub-solar-temperature difference, $\Delta T$.
The sub-solar temperature of a spherical comet nucleus is
\begin{equation}
  T = \left[\frac{(1 - A) * F_\sun}{r_h^2 \eta \epsilon
      \sigma_{SB}}\right]^{1/4},
  \label{eq:tss}
\end{equation}
where $A$ is the Bond albedo, $F_\sun = 1365$~W~m~$^{-2}$ is the solar
constant at 1~AU, $r_h$ is the heliocentric distance in AU, $\eta =
1.03$ is the IR-beaming parameter, $\epsilon = 0.95$ is the IR
emissivity of the surface, and $\sigma_{SB}$ is the Stefan-Boltzmann
constant in W~m$^{-2}$~K$^{-4}$ \citep{harris98, fernandez11}.  For
our purposes, we compute $\Delta T|_{x}$ using the minimum and maximum
$q$ values from our $\Delta q|_{x}$ integrations; $\Delta T > 0$
signifies the temperature was cooler in the past.

Figure~\ref{fig:deltaq} shows that active comets in SEPPCoN appear
more likely to have $\Delta q|_{150} < 0$ and $\Delta T|_{150} > 0$
than the inactive comets.  In addition, no active comet has $\Delta
q|_{150} > 0.5$~AU, yet $\approx10$\% of the inactive comets do have
such large values.  The K-S probabilities for $\Delta q|_{150}$ and
$\Delta T|_{150}$ are 28\% and 10\%, indicating that the difference
between active and inactive comets may be significant.  But, given the
striking differences in Fig.~\ref{fig:deltaq}, this problem deserves
future study because it may lead to an understanding of the timescales
of mantle formation.

\begin{figure}
  \ifincludegraphics
  \begin{center}
    \includegraphics[width=\columnwidth]{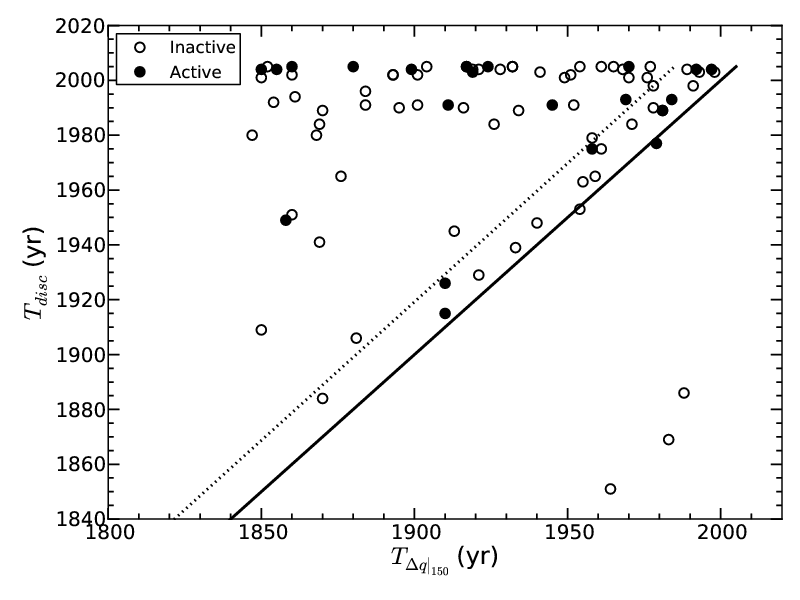}
  \end{center}
  \fi
  \caption{Year of comet discovery ($T_{disc}$) versus the year of the
    comet's 150-year minimum in perihelion distance history
    ($T_{\Delta q|_{150}}$).  Comets between the dotted and solid
    trend lines are those that were discovered near a 150-year minimum
    in perihelion distance: 8 out of the 22 comets in this region are
    active in our survey. \label{fig:discovery-and-deltaq}}
\end{figure}

The difference in perihelion distance histories between the active and
inactive comets may be related to their discovery circumstances.  When
an unknown comet is perturbed to a smaller perihelion distance, it
should become brighter to observers on the Earth, and therefore more
likely to be discovered.  To investigate this possible effect, we
searched the NASA JPL Small-Body Database Browser and the NASA
Planetary Data System Small Bodies Node for each comet's year of
discovery.  We plot this year versus the year of the minimum
$\Delta q|_{150}$ in Fig.~\ref{fig:discovery-and-deltaq}.  Data points near a
slope of 1 indicate a comet that was discovered at a 150-year low in
its perihelion distance history.  Fifteen comets are found within 20
years of this line, indicating that reductions in perihelion distance
is a factor in JFC discovery dates.  In the plot we indicate which
comets were active in our survey.  Those comets discovered soon after
being scattered to smaller perihelion distances appear to have a
higher activity rate in our survey, 8 out of 22 ($36\pm13\%$) versus
15 out of 70 ($21\pm6$\%), but the low significance precludes us from
making a firm conclusion.

Most of our survey targets were discovered in the past 20 years, but
the year of minimum $\Delta q|_{150}$ for these comets is uniformly
distributed over the past 150 years.  This distribution may be the
result of uncertain orbital histories for these comets, rather than a
lack of correlation between the two parameters.  However, it is also
possible that the increasing sophistication and sensitivity of Solar
System surveys in the past 20 years has led to JFC discoveries that
are independent of their perihelion distance history.  Both of these
possibilities could be a cause for the low significance of the K-S
tests based on $\Delta q$.

\subsubsection{Semi-major axis and \spitzer{}-comet distance}%
\label{sec:a-delta}
On average, the active comets have larger semi-major axes than the
inactive comets.  Since perihelion distance and semi-major axis are
correlated in the Jupiter-family comet population, this difference may
be an observational bias related to the perihelion distance bias
described above.  A similar argument may be made for the slight
differences in $\Delta_S$ for the active and inactive comets.  Nuclei
with larger $q$ are more likely to be observed at larger $\Delta_S$.

\subsubsection{Other investigations}\label{sec:other}
Comet lightcurve asymmetries have been recognized for many years.
Asymmetries near perihelion are not directly relevant to our survey
since most of our targets are observed at true anomalies greater than
90\degr{} from perihelion.  The secular light curves of 27 comets were
studied by \citet{ferrin10}.  Considering only the Jupiter-family
comets, and excluding the one-time active Comet 107P/Wilson-Harrington
(leaving 17 comets), the ratios of the heliocentric distance at which
activity ceases ($r_{off}$) to the distance at which activity
initiates ($r_{on}$) ranges from 0.7 to 2.0, with a mean
$r_{off}/r_{on}=1.20\pm0.38$.  Although this result cannot be directly
compared to our survey results, they qualitatively agree in that it
appears persistent post-perihelion activity is common to individual
Jupiter-family comets.

The activity of short-period comets has also been examined by
\citet{mazzottaepifani09-short}.  They searched the literature for
optical and infrared observations of short-period comets
(Jupiter-family, Encke type, and Halley type) at $r_h>3$~AU.  In their
compilation of 90 comets they find that 10--20\% are active beyond
4.5~AU.  Furthermore, they also find a perihelion asymmetry with
activity detected in 22\% of the pre-perihelion observations, and 59\%
of the post-perihelion observations.  In our survey data ($r_h>3.4$)
we find slightly different values: $24\pm5$\% (21 comets) of our
sample was active, with activity in $5\pm4$\% (2/39) of the
pre-perihelion observations, and $38\pm9$\% (19/50) in the
post-perihelion observations.  With only a few comets in each bin, and
understanding that the \citeauthor{mazzottaepifani09-short} study is
based on a literature search, and not on a consistent data set, we
consider the SEPPCoN results to be compatible with the results of
\citeauthor{mazzottaepifani09-short}.  In addition, they also conclude
that recent perihelion distance changes are not a strong predictor of
distant activity, agreeing with our results above.

\subsection{Dust color temperature}\label{sec:temperature}
The effective temperatures of comet comae at 5--20~\micron{} typically
range from $\approx0-30$\% warmer than an isothermal blackbody sphere
at the same heliocentric distance \citep{gehrz92, lisse98, sitko04,
  woodward11}.  In Table~\ref{tab:color} we find a similar result,
even though our observations are obtained at larger heliocentric
distances and longer wavelengths than are typically measured with
ground-based observatories.

We test for correlations between $T_c/T_{BB}$ for the 11 IRS center
aperture measurements and the parameters $r_h$, $q$, $F_{22}$ (22- or
24-\micron{} flux), and \efrho{}.  The results are presented in
Table~\ref{tab:efrho}.  For each parameter, we calculate the linear
correlation coefficient and apply a Spearman ``$\rho$'' test.  The
Spearman test is a more robust estimate of correlation than the linear
correlation coefficient.  It is based on the relative rankings of each
parameter being tested, and does not test for a specific functional
form of the correlation.  After computing the Spearman $\rho$, we
derive $Z$, the number of standard deviations by which the Spearman
statistic deviates from the null hypothesis expectation value for
uncorrelated data.  To account for measurement uncertainties in the
correlation tests, we tested $10^4$ data sets generated with a Monte
Carlo technique based on our measurements and associated
uncertainties.  No significant correlation is found, and the
correlations in our Monte Carlo runs never exceeded a $3\sigma$
significance.  We list our results in Table~\ref{tab:tc-tests}.

\section{Conclusions}\label{sec:conclusions}

We presented an analysis of the dust and activity of comets observed
by the \sst{} as part of the Survey of the Ensemble Physical
Properties of Cometary Nuclei (SEPPCoN).  We have shown that SEPPCoN
can be used to study the dust and activity of Jupiter-family comets at
3--7~AU, although some care must be taken when interpreting the
results as the dust detections are limited by the image sensitivities,
and that comets with large perihelion distances are a priori more
active.  We detected dust around 33 of 89 ($37\pm6$\%) survey targets,
and 21 targets ($24\pm5$\%) have comae or tails that suggest recent
cometary activity.  Since faint dust ($\tau \lesssim 0.5\ten{-9}$) may
remain undetected in our images, we conclude our detections are a
lower limit and that at least $\approx24$\% of Jupiter-family comets
are active at 3--7~AU from the Sun.  We also studied the crude
spectral properties of the dust.  The 16- to 22-\micron{} color
temperature of the dust is $7.4 \pm0.6$\% warmer on average than an
isothermal blackbody sphere in LTE.

We introduce the quantity \efrho{}, intended to be a thermal emission
counterpart to the often reported \afrho{} for observations of light
scattered by comets.  The \efrho{} versus perihelion distance
distribution of our survey sample shows that low-activity comets with
large perihelion distances are missing from the survey, and therefore
are likely missing from the known Jupiter-family comet population.

We compared the frequency of activity of survey targets to their
orbital and observational parameters.  All 30 comets with $q < 1.8$~AU
appear inactive in our survey.  Whatever drives their activities is
effectively quenched at larger distances from the Sun.

Comets that have been perturbed to smaller perihelion distances in the
past 150 years seem more likely to be active in our survey
\citep[cf.][]{licandro00}.  Kolmogorov-Smirnov tests indicate a
tentative result, so a better statistical characterization, perhaps
with more detailed orbital histories, is warranted.

The activity of Jupiter-family comets at 3--7~AU from the Sun is
significantly biased to post-perihelion epochs.  Of the 21 comets that
appear to be recently active, 19 were observed post-perihelion.  We
tentatively estimate the true anomaly at which activity is initiated
to be $f>-120$\degr, but small number statistics make this limit
uncertain.  With better confidence, we find that for many comets
activity has ceased by $f\approx 140$\degr.  Similar findings on the
persistent post-perihelion activity were found by
\citet{mazzottaepifani09-short}, and \citet{ferrin10}.  No comet in
our survey is active between $f=-180$\degr{} and $-125$\degr.

We discussed several causes for a pre-/post-perihelion activity
asymmetry.  Whatever the cause, the effect is clear in our survey.
Because SEPPCoN took a consistent approach to its observations, it can
be used to test the drivers and inhibitors of activity in a
statistical manner.  The \rosetta{} spacecraft, which will orbit Comet
67P/Churyumov-Gerasimenko \citep{glassmeier07} as it approaches
perihelion from 4~AU, should also provide new perspectives on this
problem of the evolutionary history of comet surfaces.

\section*{Acknowledgments}
The authors appreciate Paul Weissman's careful review of our
manuscript.  This work is based on observations made with the \sst,
which is operated by the Jet Propulsion Laboratory, California
Institute of Technology under a contract with NASA.  Support for this
work was, in part, provided by NASA through an award issued by
JPL/Caltech.  CS has received funding from the European Union Seventh
Framework Programme (FP7/2007-2013) under grant agreement no. 268421.

\bibliographystyle{icarus}
\bibliography{references,../drafts/extra,submitted}

\clearpage
\onecolumn
\newcommand\header{%
    \hline \\[-0.5em]
    \multicolumn{1}{c}{Name}
    & \multicolumn{1}{c}{Obs. date}
    & \multicolumn{1}{c}{$r_h$}
    & \multicolumn{1}{c}{$\Delta_S$}
    & \multicolumn{1}{c}{$\phi_S$}
    & \multicolumn{1}{c}{$f$}
    & \multicolumn{1}{c}{$T-T_p$}
    & \multicolumn{1}{c}{$a$}
    & \multicolumn{1}{c}{$e$}
    & \multicolumn{1}{c}{$q$}
    & \multicolumn{1}{c}{$\Delta q|_{150}$} \\

    & \multicolumn{1}{c}{(UT)}
    & \multicolumn{1}{c}{(AU)}
    & \multicolumn{1}{c}{(AU)}
    & \multicolumn{1}{c}{(\degr)}
    & \multicolumn{1}{c}{(\degr)}
    & \multicolumn{1}{c}{(days)}
    & \multicolumn{1}{c}{(AU)}
    & 
    & \multicolumn{1}{c}{(AU)}
    & \multicolumn{1}{c}{(AU)} \\
    \hline \\[-0.5em]
}

\setlength\LTcapwidth{6in}
\begin{scriptsize}
  \begin{longtable}[c]{lrrrrrrrrrr}
    \caption{SEPPCoN target list, observational geometry, and orbital
      parameters:
      Obs. date -- \spitzer{} observation date, for MIPS observations,
      only one of the two epochs are listed;
      $r_h$ -- heliocentric distance;
      $\Delta_S$ -- \spitzer{} comet distance;
      $\phi_S$ -- Sun-comet-\spitzer{} angle;
      $f$ -- true anomaly, 0\degr{} at perihelion;
      $T-T_p$ -- time from perihelion, $<0$ when $f<0$;
      $a$ -- semi-major axis;
      $e$ -- eccentricity;
      $q$ -- perihelion distance;
      $\Delta q|_{150}$ -- change in $q$ over the past 150~yr.
      \label{tab:targets}}\\
    \header{}
    \endfirsthead
    
    \caption{\emph{continued}}\\
    \header{}
    \endhead

    \hline
    \multicolumn{11}{l}{{\emph{Continued on next page}}} \\
    \endfoot

    \hline
    \endlastfoot

6P/d'Arrest                              \dotfill & 2007-02-13 22:48     & 4.39 & 3.85 &  12 & -145 &  -548 &  3.49 & 0.61 & 1.35 & -0.23 \\
7P/Pons-Winnecke                         \dotfill & 2007-04-19 14:39     & 4.34 & 4.03 &  13 & -146 &  -526 &  3.43 & 0.63 & 1.25 &  0.50 \\
11P/Tempel-Swift-LINEAR                  \dotfill & 2006-09-01 15:12     & 4.40 & 4.04 &  13 & -146 &  -611 &  3.42 & 0.54 & 1.56 &  0.56 \\
14P/Wolf                                 \dotfill & 2007-03-20 17:52     & 4.52 & 4.53 &  13 & -120 &  -709 &  4.25 & 0.36 & 2.72 &  1.25 \\
15P/Finlay                               \dotfill & 2007-02-28 03:06     & 4.40 & 4.32 &  13 & -149 &  -480 &  3.48 & 0.72 & 0.97 &  0.21 \\
16P/Brooks 2                             \dotfill & 2007-04-10 09:32     & 3.40 & 3.37 &  17 & -125 &  -368 &  3.36 & 0.56 & 1.47 &  0.43 \\
22P/Kopff                                \dotfill & 2007-04-19 14:10     & 4.87 & 4.38 &  11 & -156 &  -767 &  3.46 & 0.54 & 1.58 & -0.57 \\
31P/Schwassmann-Wachmann 2               \dotfill & 2006-11-09 10:30     & 5.02 & 4.59 &  11 & -167 & -1423 &  4.23 & 0.19 & 3.42 &  1.57 \\
32P/Comas Sol\`a                         \dotfill & 2006-08-01 03:59     & 4.14 & 3.76 &  14 &  122 &   486 &  4.27 & 0.57 & 1.83 & -0.42 \\
33P/Daniel                               \dotfill & 2006-11-09 17:14     & 4.40 & 4.17 &  13 & -127 &  -619 &  4.03 & 0.46 & 2.17 &  0.78 \\
37P/Forbes                               \dotfill & 2007-03-12 00:56     & 4.31 & 4.19 &  13 &  144 &   587 &  3.43 & 0.54 & 1.57 & -0.19 \\
47P/Ashbrook-Jackson                     \dotfill & 2007-03-21 10:56     & 4.32 & 4.32 &  13 & -117 &  -682 &  4.12 & 0.32 & 2.79 & -1.76 \\
48P/Johnson                              \dotfill & 2006-11-11 18:43     & 4.41 & 4.32 &  13 &  141 &   761 &  3.65 & 0.37 & 2.31 & -0.35 \\
50P/Arend                                \dotfill & 2006-10-23 18:43     & 3.48 & 3.31 &  17 & -107 &  -373 &  4.09 & 0.53 & 1.92 & -0.15 \\
51P-A/Harrington                         \dotfill & 2007-04-06 23:24     & 3.77 & 3.35 &  15 & -125 &  -438 &  3.70 & 0.55 & 1.68 &  0.19 \\
56P/Slaughter-Burnham                    \dotfill & 2006-12-23 13:48     & 5.08 & 5.04 &  12 &  120 &   708 &  5.10 & 0.50 & 2.53 &  0.12 \\
57P-A/du Toit-Neujmin-Delaporte          \dotfill & 2007-06-15 08:34     & 4.11 & 3.92 &  14 & -138 &  -560 &  3.45 & 0.50 & 1.72 &  0.46 \\
62P/Tsuchinshan 1                        \dotfill & 2006-10-03 02:00     & 4.72 & 4.66 &  12 &  150 &   664 &  3.52 & 0.58 & 1.48 & -0.84 \\
68P/Klemola                              \dotfill & 2007-02-12 09:42     & 5.47 & 5.28 &  10 & -138 &  -709 &  4.89 & 0.64 & 1.76 & -0.29 \\
69P/Taylor                               \dotfill & 2006-07-29 05:19     & 4.25 & 3.66 &  12 &  135 &   605 &  3.65 & 0.47 & 1.94 &  0.42 \\
74P/Smirnova-Chernykh                    \dotfill & 2007-01-28 17:01     & 4.42 & 4.01 &  12 & -121 &  -914 &  4.17 & 0.15 & 3.56 & -2.14 \\
77P/Longmore                             \dotfill & 2007-02-10 07:38     & 4.59 & 4.18 &  12 & -152 &  -878 &  3.60 & 0.36 & 2.31 & -0.90 \\
78P/Gehrels 2                            \dotfill & 2007-03-10 02:00     & 4.98 & 4.46 &  10 &  153 &   865 &  3.74 & 0.46 & 2.01 & -1.01 \\
79P/du Toit-Hartley                      \dotfill & 2006-09-17 23:10     & 4.37 & 4.08 &  13 & -158 &  -618 &  3.03 & 0.59 & 1.23 & -0.22 \\
89P/Russell 2                            \dotfill & 2007-05-01 09:31     & 4.76 & 4.18 &  11 & -146 &  -839 &  3.80 & 0.40 & 2.28 & -0.30 \\
93P/Lovas 1                              \dotfill & 2006-10-03 23:46     & 4.01 & 3.56 &  14 & -121 &  -439 &  4.39 & 0.61 & 1.71 &  0.07 \\
94P/Russell 4                            \dotfill & 2006-11-14 19:47     & 4.79 & 4.20 &  11 &  178 &  1174 &  3.51 & 0.36 & 2.24 & -0.43 \\
101P-A/Chernykh                          \dotfill & 2007-04-22 13:51     & 4.34 & 3.76 &  12 &  103 &   483 &  5.78 & 0.59 & 2.35 & -0.31 \\
107P/Wilson-Harrington                   \dotfill & 2007-02-12 07:43     & 4.08 & 3.52 &  13 &  166 &   582 &  2.64 & 0.62 & 0.99 &  0.05 \\
113P/Spitaler                            \dotfill & 2006-10-22 15:58     & 3.86 & 3.55 &  15 & -120 &  -518 &  3.69 & 0.42 & 2.13 &  0.36 \\
118P/Shoemaker-Levy 4                    \dotfill & 2006-09-17 11:45     & 4.95 & 4.58 &  12 &  178 &  1159 &  3.47 & 0.42 & 2.00 & -0.21 \\
119P/Parker-Hartley                      \dotfill & 2007-02-12 08:01     & 4.21 & 3.67 &  12 &  103 &   629 &  4.29 & 0.29 & 3.04 & -2.24 \\
121P/Shoemaker-Holt 2                    \dotfill & 2006-08-01 02:40     & 4.35 & 3.97 &  13 &  123 &   696 &  4.02 & 0.34 & 2.65 & -1.10 \\
123P/West-Hartley                        \dotfill & 2006-10-21 23:31     & 5.34 & 4.91 &  10 &  161 &  1049 &  3.86 & 0.45 & 2.13 & -0.23 \\
124P/Mrkos                               \dotfill & 2006-09-16 19:19     & 4.23 & 3.76 &  13 & -149 &  -588 &  3.21 & 0.54 & 1.47 & -0.45 \\
127P/Holt-Olmstead                       \dotfill & 2007-03-08 16:37     & 4.59 & 4.43 &  13 & -164 &  -957 &  3.44 & 0.37 & 2.18 & -0.21 \\
129P/Shoemaker-Levy 3                    \dotfill & 2007-04-19 00:39     & 4.01 & 3.67 &  14 &  119 &   679 &  3.76 & 0.25 & 2.81 & -3.06 \\
130P/McNaught-Hughes                     \dotfill & 2006-11-14 10:57     & 4.45 & 4.20 &  13 &  146 &   752 &  3.54 & 0.41 & 2.10 & -0.21 \\
131P/Mueller 2                           \dotfill & 2007-02-12 08:54     & 4.42 & 3.96 &  12 &  140 &   787 &  3.68 & 0.34 & 2.42 & -1.24 \\
132P/Helin-Roman-Alu 2                   \dotfill & 2007-06-07 19:26     & 3.99 & 3.80 &  15 &  120 &   478 &  4.09 & 0.53 & 1.92 &  0.28 \\
137P/Shoemaker-Levy 2                    \dotfill & 2007-03-16 13:11     & 5.47 & 5.45 &  11 & -142 &  -788 &  4.48 & 0.58 & 1.89 &  0.21 \\
138P/Shoemaker-Levy 7                    \dotfill & 2007-01-22 19:49     & 4.21 & 4.21 &  14 &  136 &   552 &  3.63 & 0.53 & 1.71 &  0.14 \\
139P/V\"ais\"al\"a-Oterma                \dotfill & 2006-11-02 07:11     & 4.10 & 3.57 &  13 &  -82 &  -535 &  4.51 & 0.25 & 3.40 & -0.83 \\
143P/Kowal-Mrkos                         \dotfill & 2007-03-09 16:14     & 4.99 & 4.74 &  11 & -133 &  -825 &  4.30 & 0.41 & 2.54 &  0.65 \\
144P/Kushida                             \dotfill & 2007-07-09 22:38     & 4.61 & 4.16 &  12 & -142 &  -567 &  3.86 & 0.63 & 1.44 & -0.11 \\
146P/Shoemaker-LINEAR                    \dotfill & 2006-08-06 20:20     & 5.08 & 4.86 &  12 & -147 &  -649 &  4.03 & 0.66 & 1.39 &  0.15 \\
148P/Anderson-LINEAR                     \dotfill & 2006-11-02 10:30     & 4.31 & 3.89 &  13 & -137 &  -567 &  3.68 & 0.54 & 1.70 & -1.36 \\
149P/Mueller 4                           \dotfill & 2007-02-09 03:36     & 5.55 & 5.46 &  10 & -150 & -1106 &  4.33 & 0.39 & 2.65 & -0.70 \\
152P/Helin-Lawrence                      \dotfill & 2006-09-17 20:19     & 5.70 & 5.58 &  10 &  158 &  1364 &  4.50 & 0.31 & 3.11 & -1.28 \\
159P/LONEOS                              \dotfill & 2007-02-13 08:58     & 6.12 & 5.74 &   9 &  118 &  1078 &  5.88 & 0.38 & 3.65 & -0.14 \\
160P/LINEAR                              \dotfill & 2006-12-19 16:49     & 4.99 & 4.83 &  12 &  144 &   798 &  3.98 & 0.48 & 2.08 & -0.68 \\
162P/Siding Spring                       \dotfill & 2007-03-17 23:43     & 4.82 & 4.27 &  11 &  173 &   858 &  3.05 & 0.60 & 1.23 &  0.14 \\
163P/NEAT                                \dotfill & 2006-08-06 17:59     & 4.09 & 3.95 &  14 &  130 &   552 &  3.67 & 0.48 & 1.92 &  0.27 \\
168P/Hergenrother                        \dotfill & 2007-05-23 02:26     & 4.49 & 4.09 &  12 &  144 &   567 &  3.63 & 0.61 & 1.42 &  1.17 \\
169P/NEAT                                \dotfill & 2007-03-01 18:40     & 4.29 & 4.01 &  13 &  168 &   530 &  2.60 & 0.77 & 0.61 & -0.05 \\
171P/Spahr                               \dotfill & 2007-03-16 13:00     & 4.16 & 4.09 &  14 &  137 &   559 &  3.53 & 0.51 & 1.73 & -0.84 \\
172P/Yeung                               \dotfill & 2006-10-18 20:36     & 4.25 & 4.06 &  14 & -141 &  -725 &  3.51 & 0.36 & 2.24 & -1.09 \\
173P/Mueller 5                           \dotfill & 2006-10-24 09:18     & 4.82 & 4.32 &  11 &  -67 &  -572 &  5.71 & 0.26 & 4.21 &  0.09 \\
197P/LINEAR                              \dotfill & 2006-11-04 17:12     & 4.21 & 4.02 &  14 & -159 &  -561 &  2.87 & 0.63 & 1.06 & -0.15 \\
213P/2005 R2 (Van Ness)                  \dotfill & 2006-11-13 12:57     & 4.05 & 3.71 &  14 &  137 &   641 &  3.43 & 0.38 & 2.13 & -0.22 \\
215P/2002 O8 (NEAT)                      \dotfill & 2007-01-08 17:45     & 4.77 & 4.24 &  11 & -161 & -1245 &  4.03 & 0.20 & 3.22 & -1.01 \\
216P/2001 CV$_8$ (LINEAR)                \dotfill & 2007-01-01 12:48     & 4.41 & 3.95 &  12 & -131 &  -648 &  3.89 & 0.44 & 2.16 & -0.43 \\
219P/2002 LZ$_{11}$ (LINEAR)             \dotfill & 2007-06-07 04:31     & 4.78 & 4.22 &  11 & -160 & -1004 &  3.65 & 0.35 & 2.37 &  0.12 \\
221P/2002 JN$_{16}$ (LINEAR)             \dotfill & 2007-04-10 02:54     & 4.40 & 3.83 &  12 & -145 &  -656 &  3.48 & 0.49 & 1.79 & -0.31 \\
223P/2002 S1 (Skiff)                     \dotfill & 2007-04-06 00:18     & 5.70 & 5.15 &   9 & -163 & -1229 &  4.14 & 0.42 & 2.41 & -0.74 \\
228P/2001 YX$_{127}$ (LINEAR)            \dotfill & 2006-10-22 11:26     & 4.84 & 4.44 &  12 &  162 &  1330 &  4.16 & 0.18 & 3.43 & -0.78 \\
246P/2004 F3 (NEAT)                      \dotfill & 2006-12-17 23:20     & 4.28 & 3.71 &  12 &  119 &   712 &  4.02 & 0.29 & 2.87 & -1.01 \\
256P/2003 HT15 (LINEAR)                  \dotfill & 2006-12-06 12:30     & 6.21 & 5.66 &  -8 & 158  &  1328 &  4.61 & 0.42 & 2.68 & -0.22 \\
260P/2005 K3 (McNaught)                  \dotfill & 2007-01-01 07:44     & 4.20 & 3.94 &  14 &  136 &   508 &  3.69 & 0.59 & 1.51 &  0.20 \\
C/2005 W2 (Christensen)                  \dotfill & 2007-01-01 04:24     & 4.07 & 3.76 &  14 &   53 &   280 & 18.93 & 0.82 & 3.33 & -0.02 \\
P/2001 R6 (LINEAR-Skiff)                 \dotfill & 2007-04-05 14:06     & 5.74 & 5.18 &   9 & -158 & -1086 &  4.15 & 0.48 & 2.17 & -0.29 \\
P/2003 O3 (LINEAR)                       \dotfill & 2007-07-09 12:00     & 4.28 & 4.10 &  14 & -153 &  -571 &  3.10 & 0.60 & 1.25 & -0.60 \\
P/2004 A1 (LONEOS)                       \dotfill & 2007-03-05 01:21     & 6.52 & 6.42 &   9 &   71 &   910 &  7.96 & 0.31 & 5.48 & -4.38 \\
P/2004 DO$_{29}$ (Spacewatch-LINEAR)     \dotfill & 2006-09-18 04:24     & 5.55 & 5.21 &  10 &   82 &   712 &  7.36 & 0.44 & 4.09 & -0.14 \\
P/2004 H2 (Larsen)                       \dotfill & 2007-01-01 20:40     & 5.46 & 5.07 &  10 &  140 &   965 &  4.51 & 0.42 & 2.62 &  0.23 \\
P/2004 V3 (Siding Spring)                \dotfill & 2006-09-18 05:33     & 5.36 & 4.90 &  10 &   82 &   676 &  7.12 & 0.45 & 3.94 & -0.20 \\
P/2004 V5-A (LINEAR-Hill)                \dotfill & 2007-03-10 01:13     & 5.81 & 5.57 &  10 &   77 &   737 &  7.98 & 0.45 & 4.42 & -0.03 \\
P/2004 VR$_8$ (LONEOS)                   \dotfill & 2006-08-07 18:51     & 3.44 & 3.35 &  17 &   85 &   339 &  4.86 & 0.51 & 2.38 & -0.19 \\
P/2005 GF$_8$ (LONEOS)                   \dotfill & 2006-11-12 23:27     & 4.16 & 3.79 &  14 &   87 &   452 &  5.85 & 0.52 & 2.83 &  0.24 \\
P/2005 JD$_{108}$ (Catalina-NEAT)        \dotfill & 2007-01-01 21:20     & 4.77 & 4.32 &  11 &   65 &   509 &  6.45 & 0.38 & 4.03 & -0.90 \\
P/2005 JQ$_5$ (Catalina)                 \dotfill & 2006-11-21 13:58     & 4.02 & 4.00 &  15 &  160 &   482 &  2.69 & 0.69 & 0.83 & -0.06 \\
P/2005 L4 (Christensen)                  \dotfill & 2007-03-11 23:02     & 4.15 & 4.07 &  14 &  116 &   564 &  4.12 & 0.43 & 2.37 &  0.76 \\
P/2005 Q4 (LINEAR)                       \dotfill & 2007-02-28 09:05     & 4.41 & 3.93 &  12 &  127 &   518 &  4.46 & 0.61 & 1.75 & -0.12 \\
P/2005 R1 (NEAT)                         \dotfill & 2006-12-19 12:41     & 4.12 & 3.72 &  14 &  108 &  -276 &  5.50 & 0.63 & 2.05 &  0.16 \\
P/2005 S3 (Read)                         \dotfill & 2007-05-23 02:48     & 4.12 & 3.88 &  14 &   93 &   500 &  4.90 & 0.42 & 2.84 & -0.15 \\
P/2005 T5 (Broughton)                    \dotfill & 2006-10-09 13:43     & 3.99 & 3.71 &  15 &   62 &   340 &  7.26 & 0.55 & 3.25 & -0.09 \\
P/2005 W3 (Kowalski)                     \dotfill & 2006-12-08 04:36     & 4.36 & 4.35 &  13 &   84 &   472 &  6.40 & 0.53 & 3.01 & -0.24 \\
P/2005 XA$_{54}$ (LONEOS-Hill)           \dotfill & 2007-06-04 23:12     & 4.42 & 3.86 &  12 &  116 &   454 &  6.13 & 0.71 & 1.77 &  0.11 \\
P/2005 Y2 (McNaught)                     \dotfill & 2007-01-20 05:02     & 5.41 & 5.07 &  10 &  101 &   753 &  6.30 & 0.47 & 3.36 &  0.14 \\

  \end{longtable}
\end{scriptsize}

{\renewcommand{\baselinestretch}{0.7}
\begin{landscape}
  \begin{table}
    \footnotesize
    \caption{Dust morphology and photometry.  Table columns:
      Instrument -- the name of the instrument used; Morph. --
      Morphology ``N'' for no dust detected, ``C'' for coma, ``Ta''
      for tail, ``Tr'' for trail, a ``?'' indicates a tentative
      detection of dust; $\rho$ -- the radius of the circular aperture
      centered on the comet nucleus; For MIPS, $F_1$ and $F_2$ are the
      flux densities of the two 23.7-\micron{} observations; For IRS,
      $F_1$ and $F_2$ are the flux desnities of the 15.8-\micron{} and
      22.3-\micron{} peak up arrays; $d$ -- The distance of the
      tail/trail aperture from the center of the comet nucleus; $A$ --
      The area of the aperture used to measure the tail/trail
      flux. \label{tab:phot}}

    \begin{tabular}{lcc crrrr @{\extracolsep{4mm}}r @{\hspace{3mm}\extracolsep{0mm}}rrrrr}
      \hline \\[-0.5em]
      & 
      & 
      & \multicolumn{5}{c}{Center aperture}

      & \multicolumn{6}{c}{Tail or trail aperture}

      \\\cline{4-8}\cline{9-14} \\[-0.5em]
      Comet
      & Instru-
      & Morph.
      & \multicolumn{1}{c}{$\rho$}
      & \multicolumn{1}{c}{$F_{1}$}
      & \multicolumn{1}{c}{$\sigma_{1}$}
      & \multicolumn{1}{c}{$F_{2}$}
      & \multicolumn{1}{c}{$\sigma_{2}$}
      & \multicolumn{1}{c}{$d$}
      & \multicolumn{1}{c}{$A$}
      & \multicolumn{1}{c}{$F_{1}$}
      & \multicolumn{1}{c}{$\sigma_{1}$}
      & \multicolumn{1}{c}{$F_{2}$}
      & \multicolumn{1}{c}{$\sigma_{2}$}
      \\
      & ment
      &
      & (px)
      & \multicolumn{2}{c}{(mJy)}
      
      & \multicolumn{2}{c}{(mJy)}
      
      & \multicolumn{1}{c}{(px)}
      & (px)
      & \multicolumn{2}{c}{(mJy)}
      
      & \multicolumn{2}{c}{(mJy)}
      
      \\
      \hline \\[-0.5em]
      
6P/d'Arrest\dotfill              & IRS & Tr?   &       6 & \nodata & \nodata & \nodata & \nodata
                                       & \nodata & \nodata & \nodata & \nodata & \nodata & \nodata  \\
16P/Brooks 2\dotfill             & MIPS  & Ta/Tr &       6 &   0.528 &   0.062 & \nodata & \nodata 
                                         & 5     &    16.9 &   0.180 &   0.023 & \nodata & \nodata  \\
22P/Kopff\dotfill                & IRS & Tr    &       6 &   0.726 &   0.075 &   1.16  &   0.17 
                                         & 11    &    91.8 &   1.109 &   0.067 &   2.53  &   0.16   \\
32P/Comas Sol\`a\dotfill         & IRS & C+Ta  &       6 &  16.21  &   0.27  &  33.10  &   0.55  
                                         & 5     &   138.9 &   8.21  &   0.30  &  19.15  &   0.62   \\
37P/Forbes\dotfill               & IRS & Ta/Tr &       6 & \nodata & \nodata &   0.625 &   0.051 
                                         & 7     &    58.3 &   0.205 &   0.028 &   0.536 &   0.036  \\
48P/Johnson\dotfill              & IRS & Ta    &       6 &   0.61  &   0.16  &   0.69  &   0.21 
                                         & 12    &    47.9 &   0.63  &   0.10  &   0.99  &   0.13   \\
50P/Arend\dotfill                & IRS & Ta?   & \nodata & \nodata & \nodata & \nodata & \nodata
                                       & \nodata & \nodata & \nodata & \nodata & \nodata & \nodata  \\
56P/Slaughter-Burnham\dotfill    & IRS & Ta/Tr &       3 & \nodata & \nodata & \nodata & \nodata 
                                         & 15    &    77.2 &   0.216 &   0.027 &   0.315 &   0.058  \\
62P/Tsuchinshan 1\dotfill        & MIPS  & Tr    &       4 &   0.177 &   0.034 &   0.242 &   0.050 
                                         & 12    &    72.0 &   0.377 &   0.040 & \nodata & \nodata  \\
69P/Taylor\dotfill               & IRS & C     &       4 & \nodata & \nodata &   0.477 &   0.031 
                                       & \nodata & \nodata & \nodata & \nodata & \nodata & \nodata  \\
74P/Smirnova-Chernykh\dotfill    & IRS & C+Tr  &       6 &   6.52  &   0.19  &  13.30  &   0.19  
                                         & 14    &   104.2 &   1.55  &   0.18  &   3.28  &   0.18   \\
78P/Gehrels 2\dotfill            & IRS & Ta/Tr & \nodata & \nodata & \nodata & \nodata & \nodata 
                                         & 13    &    59.8 &   0.320 &   0.035 &   0.804 &   0.084  \\
101P-A/Chernykh\dotfill          & IRS & C     &       6 &   1.136 &   0.060 &   3.20  &   0.13 
                                       & \nodata & \nodata & \nodata & \nodata & \nodata & \nodata  \\
118P/Shoemaker-Levy 4\dotfill    & IRS & Ta    &       6 &   0.248 &   0.074 & \nodata & \nodata 
                                         & 6     &    64.9 & \nodata & \nodata &   0.405 &   0.085  \\
119P/Parker-Hartley\dotfill      & IRS & Ta    &       6 &   1.81  &   0.10  &   4.16  &   0.18  
                                         & 12    &    88.0 &   1.914 &   0.088 &   4.27  &   0.14   \\
121P/Shoemaker-Holt 2\dotfill    & IRS & Tr    & \nodata & \nodata & \nodata & \nodata & \nodata 
                                         & 13    &   148.4 &   2.33  &   0.13  &   4.98  &   0.23  \\
129P/Shoemaker-Levy 3\dotfill    & IRS & Ta    &       6 & \nodata & \nodata &   4.65  &   0.23 
                                         & 7     &    43.6 & \nodata & \nodata &   3.05  &   0.14   \\
144P/Kushida\dotfill             & MIPS  & Tr    &       6 & \nodata & \nodata & \nodata & \nodata 
                                         & 27    &   142.2 & \nodata & \nodata &   0.61  &   0.18   \\
152P/Helin-Lawrence\dotfill      & IRS & Ta    &       6 & \nodata & \nodata &   0.427 &   0.072 
                                         & 12    &    57.4 &   0.321 &   0.038 &   0.818 &   0.050  \\
159P/LONEOS\dotfill              & IRS & Ta    &       6 &   0.315 &   0.036 &   0.539 &   0.060 
                                         & 15    &    57.6 &   0.259 &   0.025 &   0.891 &   0.042  \\
171P/Spahr\dotfill               & IRS & Tr    & \nodata & \nodata & \nodata & \nodata & \nodata 
                                         & 15    &    50.2 &   0.285 &   0.030 &   0.293 &   0.038  \\
173P/Mueller 5\dotfill           & IRS & Ta    &       6 &   6.75  &   0.26  &  16.96  &   0.27  
                                         & 15    &   111.9 &   4.65  &   0.26  &  13.03  &   0.27   \\
213P/2005 R2 (Van Ness)\dotfill  & IRS & C+Ta  &       6 &  14.96  &   0.39  &  32.35  &   0.74  
                                         & 14    &   224.7 &  21.82  &   0.57  &  45.6   &   1.1   \\
219P/2002 LZ$_{11}$
                (LINEAR)\dotfill & MIPS  & Tr    &       4 & \nodata & \nodata & \nodata & \nodata 
                                         & 37    &   217.9 &   1.01  &   0.11  &   0.98  &   0.10   \\
246P/2004 F3 (NEAT)\dotfill      & IRS & C+Ta  &       6 &  25.57  &   0.48  &  59.2   &   1.1   
                                         & 11    &   327.5 &  36.33  &   0.88  &  87.6   &   2.0    \\
260P/2005 K3 (McNaught)\dotfill  & MIPS  & Tr    & \nodata & \nodata & \nodata & \nodata & \nodata 
                                         & 38    &   402.2 &   2.33  &   0.12  & \nodata & \nodata  \\
C/2005 W2 (Christensen)\dotfill  & MIPS  & Ta    &       6 & \nodata & \nodata &   1.41  &   0.14  
                                         & 24    &   163.9 &   2.04  &   0.18  &   2.68  &   0.17   \\
P/2004 A1 (LONEOS)\dotfill       & MIPS  & C+Ta  &       6 &   6.65  &   0.14  &   6.10  &   0.11  
                                         & 21    &   217.9 &   6.55  &   0.20  &   5.98  &   0.16   \\
P/2004 H2 (Larsen)\dotfill       & MIPS  & C     &       4 &   0.415 &   0.062 & \nodata & \nodata 
                                       & \nodata & \nodata & \nodata & \nodata & \nodata & \nodata  \\
P/2004 V5-A
           (LINEAR-Hill)\dotfill & IRS & Ta    &       6 &   0.640 &   0.048 &   1.463 &   0.087 
                                         & 14    &    79.3 &   0.350 &   0.040 &   0.988 &   0.072  \\
P/2004 VR$_8$ (LONEOS)\dotfill   & IRS & Ta    &       6 & \nodata & \nodata &  11.6   &   1.1
                                         & 12    &   125.5 &  14.87  &   0.79  &  28.8   &   1.2    \\
P/2005 JD$_{108}$
         (Catalina-NEAT)\dotfill & MIPS  & C+Ta  &       6 &  10.20  &   0.14  &  10.53  &   0.14  
                                         & 23    &   212.5 &   9.68  &   0.20  &  10.13  &   0.20   \\
P/2005 T5 (Broughton)\dotfill    & MIPS  & Tr    &       4 & \nodata & \nodata &   0.44  &   0.12  
                                         & 22    &   129.1 &   1.90  &   0.25  &   1.58  &   0.20   \\
P/2005 W3 (Kowalski)\dotfill     & MIPS  & Ta    &       5 &   2.30  &   0.14  &   2.27  &   0.15  
                                         & 12    &    81.9 &   1.85  &   0.14  & \nodata & \nodata  \\
P/2005 Y2 (McNaught)\dotfill     & MIPS  & C+Ta  &       6 &  13.94  &   0.22  &  15.79  &   0.23  
                                         & 17    &   173.1 &   6.79  &   0.27  &   7.37  &   0.29   \\
\hline
    \end{tabular}
  \end{table}
\end{landscape}
}

\renewcommand{\baselinestretch}{0.7}
\begin{table}
  \centering
  \caption{Color-temperature excess ($T_c/T_{BB}$) for IRS peak up
    observations with photometry in both filters
    (Table~\ref{tab:phot}).  Only temperatures with uncertainties less
    than 40~K are listed.  Table columns: $T_{BB}$, the temperature of
    an isothermal blackbody sphere in local thermodynamic equilibrium
    at the same $r_h$; Center, the color temperature and error,
    normalized by $T_{BB}$, for the center aperture of
    Table~\ref{tab:phot}; Tail or Trail, the same as Center, but for
    the tail or trail aperture.\label{tab:color}}

  \begin{tabular}{lrr cr c cr}
    \hline \\[-0.5em]
    Comet
    & \multicolumn{1}{c}{$r_h$}
    & \multicolumn{1}{c}{$T_{BB}$}
    & \multicolumn{2}{c}{Center}

    && \multicolumn{2}{c}{Tail or Trail}

    \\\cline{4-5}\cline{7-8} \\[-0.5em]
    & (AU)
    & (K)
    & \multicolumn{1}{c}{$T_c/T_{BB}$}
    & \multicolumn{1}{c}{$\sigma$}
    && \multicolumn{1}{c}{$T_c/T_{BB}$}
    & \multicolumn{1}{c}{$\sigma$}
    \\
    \hline \\[-0.5em]

22P            \dotfill & 4.87 &  126 &    1.43 &    0.23 &&    1.14 &    0.05 \\
32P            \dotfill & 4.14 &  137 &    1.12 &    0.02 &&    1.04 &    0.03 \\
37P            \dotfill & 4.31 &  134 & \nodata & \nodata &&    1.00 &    0.08 \\
48P            \dotfill & 4.41 &  132 & \nodata & \nodata &&    1.37 &    0.24 \\
74P            \dotfill & 4.42 &  132 &    1.16 &    0.02 &&    1.13 &    0.09 \\
78P            \dotfill & 4.98 &  125 & \nodata & \nodata &&    1.09 &    0.09 \\
101P           \dotfill & 4.34 &  133 &    0.96 &    0.03 && \nodata & \nodata \\
119P           \dotfill & 4.21 &  135 &    1.06 &    0.04 &&    1.07 &    0.03 \\
121P           \dotfill & 4.35 &  133 & \nodata & \nodata &&    1.12 &    0.05 \\
152P           \dotfill & 5.70 &  116 & \nodata & \nodata &&    1.16 &    0.08 \\
159P           \dotfill & 6.12 &  112 &    1.52 &    0.17 &&    1.04 &    0.05 \\
173P           \dotfill & 4.82 &  127 &    1.08 &    0.02 &&    1.02 &    0.03 \\
213P           \dotfill & 4.05 &  138 &    1.07 &    0.02 &&    1.09 &    0.02 \\
246P           \dotfill & 4.28 &  134 &    1.06 &    0.01 &&    1.04 &    0.02 \\
P/2004 V5-A    \dotfill & 5.81 &  115 &    1.24 &    0.07 &&    1.11 &    0.07 \\
P/2004 VR$_8$  \dotfill & 3.44 &  150 & \nodata & \nodata &&    1.05 &    0.04 \\

    \hline

  \end{tabular}
\end{table}
\renewcommand{\baselinestretch}{1.0}

\renewcommand{\baselinestretch}{0.7}
\begin{table}
  \centering
  \caption{Kolmogorov-Smirnov statistic, $D$, and $p$-value used to
    determine if active comets and inactive comets are drawn from
    different distributions, based on 7 parameters: $r_h$,
    heliocentric distance; $\Delta_S$, \spitzer-comet distance;
    $\phi_S$, phase angle; $f$, true anomaly; $a$, orbital semi-major
    axis; $e$, orbital eccentricity; $q$, perihelion distance; $\Delta
    q|_{150}$, change in $q$ over the past 150 years.  Active comets
    are those 21 targets that were observed to have a coma and/or
    tail.  As an additional test, we removed comets with tail-only
    morphologies from the active list, leaving 10 comets with coma
    detections.  \label{tab:kstests}}

  \begin{tabular}{rclcl}
    \hline \\[-0.5em]

    & \multicolumn{2}{c}{Active vs. Inactive}
    & \multicolumn{2}{c}{Coma vs. Inactive}
    \\
    Parameter
    & $D$
    & $p$
    & $D$
    & $p$
    \\
    \hline \\[-0.5em]

    $r_h$            & 0.18 & 0.67     & 0.19 & 0.88 \\
    $\Delta_S$       & 0.29 & 0.12     & 0.36 & 0.17 \\
    $\phi_S$         & 0.20 & 0.52     & 0.33 & 0.24 \\
    $f$              & 0.51 & 0.00024  & 0.48 & 0.025 \\
    $a$              & 0.45 & 0.0021   & 0.43 & 0.053 \\
    $e$              & 0.34 & 0.039    & 0.38 & 0.13 \\
    $q$              & 0.50 & 0.00045  & 0.44 & 0.051 \\
    $\Delta q|_{150}$ & 0.24 & 0.28     & 0.31 & 0.33 \\
    $\Delta T|_{150}$ & 0.30 & 0.10     & 0.26 & 0.52 \\
    \hline

  \end{tabular}
\end{table}
\renewcommand{\baselinestretch}{1.0}

\renewcommand{\baselinestretch}{0.7}
\begin{table}
  \centering

  \caption{\efrho{} values for all center aperture photometry in
    Table~\ref{tab:phot}.  $\efrho_{16}$ and $\efrho_{22}$ are
    computed from IRS peak-up photometry, $\efrho_{24}$ is computed
    from MIPS photometry.  For comets with two epochs of MIPS
    photometry, \efrho{} is computed from the average.  For all
    comets, $\epsilon$ is computed assuming a grain temperature 10\%
    warmer than a blackbody in LTE. \label{tab:efrho}}

  \begin{tabular}{lrrrrrr}
    \hline \\[-0.5em]
    Comet
    & $\efrho_{16}$
    & $\sigma_{16}$
    & $\efrho_{22}$
    & $\sigma_{22}$
    & $\efrho_{24}$
    & $\sigma_{24}$
    \\

    & \multicolumn{2}{c}{(cm)}
    & \multicolumn{2}{c}{(cm)}
    & \multicolumn{2}{c}{(cm)}
    \\
    \hline \\[-0.5em]
16P         \dotfill & \nodata & \nodata & \nodata & \nodata &  1.4    &  0.2 \\
22P         \dotfill & 20.1    &  2.1    & 13.2    &  2.0    & \nodata & \nodata \\
32P         \dotfill &230.3    &  3.8    &223.6    &  3.7    & \nodata & \nodata \\
37P         \dotfill & \nodata & \nodata &  5.1    &  0.4    & \nodata & \nodata \\
48P         \dotfill & 12.1    &  3.2    &  6.2    &  1.9    & \nodata & \nodata \\
62P         \dotfill & \nodata & \nodata & \nodata & \nodata &  2.3    &  0.3 \\
69P         \dotfill & \nodata & \nodata &  5.0    &  0.3    & \nodata & \nodata \\
74P         \dotfill &120.8    &  3.6    &110.7    &  1.6    & \nodata & \nodata \\
101P-A      \dotfill & 18.7    &  1.0    & 23.9    &  1.0    & \nodata & \nodata \\
118P        \dotfill &  7.6    &  2.2    & \nodata & \nodata & \nodata & \nodata \\
119P        \dotfill & 26.4    &  1.5    & 28.4    &  1.1    & \nodata & \nodata \\
129P        \dotfill & \nodata & \nodata & 28.6    &  1.4    & \nodata & \nodata \\
152P        \dotfill & \nodata & \nodata &  9.1    &  1.5    & \nodata & \nodata \\
159P        \dotfill & 25.3    &  2.9    & 14.1    &  1.6    & \nodata & \nodata \\
173P        \dotfill &178.0    &  6.8    &185.5    &  3.0    & \nodata & \nodata \\
213P        \dotfill &196.3    &  5.1    &205.6    &  4.7    & \nodata & \nodata \\
246P        \dotfill &397.0    &  7.4    &424.6    &  7.9    & \nodata & \nodata \\
C/2005 W2   \dotfill & \nodata & \nodata & \nodata & \nodata &  6.1    &  0.6 \\
P/2004 A1   \dotfill & \nodata & \nodata & \nodata & \nodata &139.1    &  2.0 \\
P/2004 H2   \dotfill & \nodata & \nodata & \nodata & \nodata &  6.9    &  1.0 \\
P/2004 V5-A \dotfill & 41.3    &  3.1    & 32.6    &  1.9    & \nodata & \nodata \\
P/2004 VR$_8$\dotfill & \nodata & \nodata & 47.5   &  4.6    & \nodata & \nodata \\
P/2005 JD$_{108}$\dotfill & \nodata & \nodata & \nodata & \nodata & 72.6 &  0.7 \\
P/2005 T5   \dotfill & \nodata & \nodata & \nodata & \nodata &  2.3    &  0.6 \\
P/2005 W3   \dotfill & \nodata & \nodata & \nodata & \nodata & 15.9    &  0.7 \\
P/2005 Y2   \dotfill & \nodata & \nodata & \nodata & \nodata &162.1    &  1.7 \\

    \hline
  \end{tabular}
\end{table}
\renewcommand{\baselinestretch}{1.0}

\renewcommand{\baselinestretch}{0.7}
\begin{table}
  \centering

  \caption{Linear correlation coefficients and Spearman $\rho$ tests
    used to test if color-temperature excess ($T/T_{BB}$) for the 11
    IRS peak-up center apertures (Table~\ref{tab:color}) is correlated
    with heliocentric distance, flux, and \efrho{} (flux and \efrho{}
    are measured at 22 or 24~\micron).  For the Spearman test: $Z$ is
    the number of standard deviations by which the Spearman $\rho$
    deviates from its null hypothesis expected value; $P_Z$ is the
    probability that a value of $Z$ or greater may occur with
    uncorrelated data sets; $\langle Z \rangle$ is the mean $Z$ from
    Monte Carlo runs based on our measurements and
    errors.\label{tab:tc-tests}}

  \begin{tabular}{rcccr}
    \hline \\[-0.5em]
    Parameter
    & $r$
    & $Z$
    & $P_Z$
    & $\langle Z \rangle$
    \\
    \hline \\[-0.5em]

    $r_h$        &  0.77 & 2.1 & 0.03 & 1.8 \\
    $q$          &  0.09 & 0.6 & 0.60 & 0.5 \\
    $F_{22}$      & -0.44 & 1.6 & 0.12 & 1.5 \\
    \efrho$_{22}$ & -0.44 & 1.1 & 0.31 & 1.1 \\

    \hline

  \end{tabular}
\end{table}
\renewcommand{\baselinestretch}{1.0}

\clearpage
\appendix
\section{The Quantity \efrho{}}\label{sec:efrho}

For observations of light scattered by comet comae, \citet{ahearn84}
defined the quantity \afrho{}, where $A$ is the grain albedo, and $f$
is the areal filling factor within an observed aperture of radius
$\rho$.  The \afrho{} parameter has been widely reported for visible
light observations of comets, and, when given in units of cm, is
crudely proportional to dust production rate in kg~s\inv{}
\citep{ahearn95}.  We have defined a new parameter, \efrho, that we
intend to be an analogous quantity for observations of thermal
emission from comet comae.  The $\epsilon$ in \efrho{} is the
effective emissivity of the dust grains; $f$ and $\rho$ remain the
same as for \afrho.  Both \afrho{} and \efrho{} are independent of
aperture size for comae with a $\rho$\inv{} mean surface brightness
profile.  To compute \efrho{} from observations, we provide the
formula
\begin{equation}
  \epsilon f \rho = \frac{I_\lambda \rho}{B_\lambda(T_c)} ,
\end{equation}
where $I_\lambda$ is the observed mean surface brightness in an
aperture of radius $\rho$ centered on the comet, $B_\lambda$ is the
Planck function evaluated at $T_c$ with same units as $I_\lambda$
(e.g., MJy~sr\inv), and $T_c$ is the temperature of the effective
continuum.  In the mid-IR, $T_c = 1.1\,T_{BB}$ can usually be assumed
in the absence of any color information, as discussed in
\S\ref{sec:temperature}.  However, it is important to note that
\efrho{} is sensitive to the adopted $T_c$, and variations by factors
of 2 in \efrho{} can easily be accounted for by small variations in
$T_c$.

To give our SEPPCoN observations context, we present \efrho{} values
based on mid-IR observations of several comets in the literature
(Table~\ref{tab:efrho}).  Continuum temperatures were taken from the
original investigations when possible; we assumed $T_c=1.1\,T_{BB}$
for the observation of Comet 2P/Encke at 2.5~AU, and $1.08\,T_{BB}$
for 103P/Hartley~2, reported by \citet{sitko11-iauc9181}.  The flux of
Comet Encke's nucleus, based on a 2.3~km effective radius
\citep{kelley06-comets}, has been subtracted from the Encke photometry
of \citet{reach07} and \citet{gehrz89} (0.04 and 1.5~Jy,
respectively).  For Comet Hartley 2, we computed a nucleus flux of
0.97~Jy assuming a 0.6~km effective radius \citep{groussin04-comets,
  lisse09}.  Nuclei were computed using the near-Earth asteroid
thermal model of \citet{harris98}, with an IR-beaming parameter
$\eta=1.0$.  The nucleus fluxes for the remaining comets are assumed
to be negligible.

\begin{landscape}\begin{table}\footnotesize
  \centering

  \caption{\efrho{} values computed from selected comet observations
    in the literature, presented for comparison to the comets in our
    survey.  The table has been sorted by
    \efrho. \label{tab:efrho-compare}}

  \begin{tabular}{rccrrrl}
    \hline \\[-0.5em]
    Comet
    & \multicolumn{1}{c}{$r_h$}
    & \multicolumn{1}{c}{$\lambda$}
    & \multicolumn{1}{c}{$\rho$}
    & \multicolumn{1}{c}{$\efrho$}
    & \multicolumn{1}{c}{$\sigma$}
    & Notes
    \\

    & \multicolumn{1}{c}{(AU)} 
    & \multicolumn{1}{c}{(\micron)}
    & \multicolumn{1}{c}{(km)}
    & \multicolumn{1}{c}{(cm)}
    & \multicolumn{1}{c}{(cm)}
    \\
    \hline \\[-0.5em]
    2P/Encke                         & 2.52 & 23.7 & 18\,300 &          31 & \nodata      & \citet{reach07} \\
    73P-B/Schwassman-Wachmann 3      & 1.11 & 11.7 &     430 &         744 &        8     & $\approx1$ week post-outburst, \citet{harker11} \\
    103P/Hartley 2                   & 1.06 & 11.6 &     330 &         814 &       26     & \citet{meech11-epoxi} \\
    73P-C/Schwassman-Wachmann 3      & 1.09 & 11.7 &     350 &      2\,518 &       25     & \citet{harker11} \\
    2P/Encke                         & 0.38 & 10.0 &  2\,000 &      2\,720 &       60     & \citet{gehrz89} \\
    C/1995 O1 (Hale-Bopp)            & 2.8  & 7.8  &  3\,300 &     52\,400 &   5\,200     & pre-perihelion, \citet{wooden99} \\
    C/1996 B2 (Hyakutake)            & 0.34 & 10.3 & 14\,400 &     53\,500 &   1\,500     & \citet{mason98} \\
    1P/Halley                        & 0.59 & 10.3 & 12\,600 &    127\,000 & \nodata      & \citet{gehrz92} \\
    C/1995 O1 (Hale-Bopp)            & 0.92 & 10.3 & 13\,400 & 1\,280\,000 & 120\,000     & \citet{mason01} \\

    \hline
  \end{tabular}
\end{table}\end{landscape}
\renewcommand{\baselinestretch}{1.0}

\end{document}